\documentclass[fleqn,usenatbib]{mnras}
\usepackage[T1]{fontenc}
\DeclareRobustCommand{\VAN}[3]{#2} \let\VANthebibliography\thebibliography
\def\thebibliography{\DeclareRobustCommand{\VAN}[3]{##3}\VANthebibliography}

\usepackage{graphicx}
\usepackage{amsmath}
\usepackage{amssymb}
\usepackage{booktabs}
\usepackage{xcolor}
\usepackage{CJK}
\usepackage{ulem}
\usepackage{newtxtext,newtxmath}  
\newcommand{\dataavailability}[1]{\begin{small}\section*{Data
Availability}\end{small}{\noindent #1}\vspace{5pt}}

\title[High Mach number radiative TMLs]{Radiative turbulent mixing layers at high Mach numbers}

\author[Y. Yang \& S. Ji]{Yanhui Yang (杨焱辉)$^{1,2}$ and Suoqing
Ji (季索清)$^{1}$\thanks{E-mail: suoqing@shao.ac.cn} \\
$^{1}$Astrophysics Division \& Key Laboratory for Research in Galaxies and
Cosmology, Shanghai Astronomical Observatory, Chinese Academy
of Sciences, \\
~80 Nandan Road, Shanghai 200030, China \\
$^{2}$Department of Astronomy, University of
Science and Technology of China, Hefei 230026, China\\}

\date{Accepted XXX. Received YYY; in original form ZZZ}

\pubyear{2022}

\begin{document}
\begin{CJK}{UTF8}{gbsn}
\label{firstpage}
\pagerange{\pageref{firstpage}--\pageref{lastpage}}
\maketitle

\begin{abstract}
   Radiative turbulent mixing layers (TMLs) are ubiquitous in astrophysical
   environments, e.g., the circumgalactic medium (CGM), and are triggered by the
   shear velocity at interfaces between different gas phases. To understand the
   shear velocity dependence of TMLs, we perform a set of 3D hydrodynamic
   simulations with an emphasis on the TML properties at high Mach numbers
   $\mathcal{M}$. Since the shear velocity in mixing regions is limited by the
   local sound speed of mixed gas, high-Mach number TMLs develop into a two-zone
   structure: a Mach number-independent mixing zone traced by significant
   cooling and mixing, plus a turbulent zone with large velocity dispersions
   which expands with greater $\mathcal{M}$. Low-Mach number TMLs do not have
   distinguishable mixing and turbulent zones. The radiative cooling of TMLs at low and high Mach numbers is predominantly balanced by enthalpy consumption and turbulent dissipation respectively. Both the TML surface brightness and column
   densities of intermediate-temperature ions (e.g., \ion{O}{VI}) scale as
   $\propto\mathcal{M}^{0.5}$ at $\mathcal{M} \lesssim 1$, but reach saturation
   ($\propto \mathcal{M}^0$) at $\mathcal{M} \gtrsim 1$. Inflow velocities and
   hot gas entrainment into TMLs are substantially suppressed at high Mach
   numbers, and strong turbulent dissipation drives the evaporation of cold gas.
   This is in contrast to low-Mach number TMLs where the inflow velocities and
   hot gas entrainment are enhanced with greater $\mathcal{M}$, and cold gas
   mass increases due to the condensation of entrained hot gas.

\end{abstract}


\section{Introduction}

Nowadays, the circumgalactic medium (CGM) has become a new frontier in studies
of the formation and evolution of galaxies. Many puzzles are still under debate
\citep{tumlinson2017circumgalactic}, among which is the ubiquity of ions peaking
at the ``warm'' temperatures of $\sim 10^5\,\mathrm{K}$, such as \ion{O}{VI},
detected in quasar absorption sightlines intersecting the CGM
\citep[e.g.,][]{Tumlinson2005,Prochaska2011,Tumlinson2011,Savage2014,Lehner2015,Johnson2015,Keeney2017,qu2020warm}.
Such ions are also discovered in high-velocity clouds (HVCs) around the Milky
Way Galaxy \citep[e.g.,][]{Savage2014,tripp2022high}. However, the origin and
sustainability of these ions are controversial, since gas enriched with
\ion{O}{VI} at $\sim 10^5\,\mathrm{K}$ is thermally unstable with a cooling time
of $\sim 10\,\mathrm{Myr}$ much shorter than the dynamic timescale. One possible
origin of \ion{O}{VI} is the mixing of cold ($\sim 10^4\,\mathrm{K}$) and hot
($\sim 10^6\,\mathrm{K}$) phases, given the existence of large amounts of both
hot \citep{li2008chandra,fang2012hot,planck13,anderson15} and cold
\citep{Stocke2013,Werk2014,Prochaska2017, cantalupo14, hennawi15, cai17} gas in
the CGM. The turbulent mixing of two phases, usually accompanied by efficient
radiative cooling, could be triggered by the Kelvin-Helmholtz instability
occurring at the cold-hot interface with shear flows, which is referred to as
so-called radiative turbulent mixing layers (TMLs).

Theories of radiative TMLs were proposed by \citet{Begelman1990} and
\citet{Slavin1993}, and numerical experiments were latter carried out by
\citet{Kwak2010}. \citet{Ji2019} confronted analytic models against 3D numerical
simulations, finding that the thickness of mixing layer $l$ is determined by
radiative cooling rather than the KH instability, and the column densities of warm ions are
not independent of density and metallicity as previously predicted, but show
weak dependence on shear velocity and density contrast between the cold and
hot phases. Unlike adiabatic TMLs which cannot reach equilibrium with a finite thickness, radiative TMLs ultimately reach a dynamic quasi-equilibrium state with a saturated TML thickness. The quasi-equilibrium state can be described by the
balance between the flux of enthalpy and kinetic energy and radiative cooling as
follows (Eq.~(22) in \citealt{Ji2019}):
\begin{equation}
    Q \sim \frac{5}{2}P v_\text{in} \left(1+\frac{1}{3}\mathcal{M}^2 \right),
    \label{eq:balance}
\end{equation}
where $Q$ is the surface brightness of TMLs, $P$ the pressure of hot gas,
$v_\text{in}$ the inflow velocity which drives hot gas entrainment, and
$\mathcal{M}$ the Mach number. \citet{Fielding2020} further argued that the thin
corrugated cooling sheet in strong cooling mixing layers has an area with
fractal dimension of $5/2$, and found that hot gas entrainment into TMLs is
enhanced by shorter cooling time, larger shear velocity and larger density
contrast. \citet{Tan2021} put forward novel insights into radiative TMLs from
the view point of turbulent combustion theories, and adopted the Damk{\"o}hler
number defined by $\text{Da}\equiv \tau_\text{turb}/t_\text{cool}$ to quantify
the strength of radiative cooling in TMLs, where $\tau_\text{turb}$ and
$t_\text{cool}$ are the outer eddy turnover time and cooling time respectively.
$\text{Da}>1$ indicates the strong cooling regime where TMLs are highly
multiphase and exhibit fractal structures \citep{Fielding2020}, whereas
$\text{Da}<1$ corresponds to TMLs in the weak cooling regime with almost
single-phase mixed gas \citep{Ji2019}. \citet{Tan2021} also explicitly
demonstrated the scaling relations of TML surface brightness $Q$, with $Q\propto
t_\text{cool}^{-1/2}$ \citep{Ji2019} and $Q\propto t_\text{cool}^{-1/4}$
\citep{Fielding2020,Mandelker2020} in the weak and strong cooling regimes
respectively. These progresses on the importance of radiative cooling in TMLs
also shed light on related topics, in particular, the survival of cold clouds in
hot winds, which could be partially viewed as a ``zoom-out'' version of TMLs \citep{Klein1994,Mellema2002,Scannapieco2015,Brueggen2016,Schneider2017,Armillotta2017}.
\citet{Gronke2018} 
found under sufficiently strong cooling, wind materials are rapidly entrained
into radiative TMLs, condense into cold phase and lead to the growth of cold
clouds
\citep{Li2019,Sparre2020,Gronke2021a,Kanjilal2020,Farber2021,banda2021shock,bustard2022radiative,abruzzo2022simple}.
More recently, different cold cloud kinematics, e.g., infalling clouds under
gravity \citep{tan2023cloudy}, are also considered.

Although the dependency of radiative TMLs on various above-mentioned
properties is under active investigation,
literatures focusing on the impact of shear velocity on TMLs are relatively
sparse, especially when the shear velocity becomes supersonic. Though
\citet{Ji2019} discussed the shear velocity dependency, the parameter space is
relatively small with the maximum shear velocities to be transonic only. In
\citet{Fielding2020} and \citet{Tan2021}, TMLs in the strong cooling regime are
studied in depth, while the shear velocity dependency is not their primary
focus. \citet{Tan2021} presented scaling relations of physical quantities
including $Q \propto u'^{1/2}$ and $Q \propto u'^{3/4}$ in the weak and strong
cooling regimes respectively with $u'$ as turbulent velocity, but it is unclear
whether these scaling relations could be safely extrapolated to supersonic
regimes. In reality, transonic and even supersonic flows are ubiquitous in
observations and recent hydrodynamic simulations, say, associated with cold
streams feeding galaxies
\citep{kerevs2005galaxies,dekel2009cold,van2011rates,faucher2011baryonic} and
subsequent virial shocks in the CGM
\citep[e.g.,][]{Birnboim2003,dekel2006galaxy,oppenheimer2010feedback,stern2021virialization,ji2021virial},
as well as AGN jets
\citep{marscher2002observational,guo2012fermi,yuan2014hot,li2015cooling,yang2016agn}.
Supersonic
galactic outflows
\citep[e.g.,][]{Veilleux1999,kim2017three,li2020supernovae,schneider2020physical,fielding2022structure,girichidis2021situ,hopkins2021cosmic,sarkar2022self}
are also capable to produce strong turbulence or velocities shears. Some
previous works on supersonic TMLs have found the suppression of the KH
instability, including both fluid mechanics experiments in early years
(e.g.,
\citealt{Chinzei1986,Papamoschou1988,Goebel1991,Clemens1992,Barre1994,NAUGHTON1997,SLESSOR2000})
and recent applications in astrophysical systems
(e.g., \citealt{Scannapieco2015,white2016turbulent,Mandelker2016}).
However, the effects of radiative cooling are either excluded or relatively less
explored. For instance, \citet{Mandelker2016} theoretically showed that modes
unstable to the KH instability are damped in the classic planar sheet case when
the Mach number $\mathcal{M}$ exceeds a critical value:
\begin{equation}
    \mathcal{M}_\text{crit} = \left(1+\delta ^{-1/3}\right)^{3/2},
\end{equation}
where $\mathcal{M}$ is with respect to the sound speed in the hot medium
$c_\text{s,h}$, and $\delta \equiv \rho_\text{cold}/\rho_\text{hot}$ is the
density contrast of two phases. But only adiabatic cases are considered in
\citet{Mandelker2016}. \citet{Mandelker2020} further incorporated radiative
cooling in cold stream simulations and found significant cooling-induced mass
growth of cold streams, while the detailed supersonic TML physics and its
quantitative velocity dependence are not the main focus. Therefore, it is still
unclear whether the suppression of the KH instability is applicable to TMLs with
radiative cooling, or how supersonic TMLs with radiative cooling will evolve in
the long run.

Given both the importance of cooling in TMLs and the ubiquity of supersonic
shear flows, it might be intriguing to extend existing studies on radiative TMLs
further to supersonic conditions, which motivates this work. We also limit the
scope of this work to the weak cooling regime, which is realistic for CGM-like
astrophysical environments where the gas pressure and metallicity are relatively
low. The outline of this paper is as follows. In \S\ref{sec:methods}, we
describe our numerical methods. We present our results in \S\ref{sec:results},
and finally discuss our findings and concludes in \S\ref{sec:diss}.

\section{Methods}
\label{sec:methods}

We use the {\small FLASH} code \citep{Fryxell2000} developed by the FLASH Center
of the University of Chicago and the University of Rochester for our
simulations, where the HLLC Riemann solver is adopted to solve the equations of
inviscid ideal hydrodynamics:
\begin{subequations}
    \begin{gather}
        \frac{\partial \rho }{\partial t} + \left(\bmath{v}\cdot
        \nabla \right)\rho + \rho \nabla \cdot \bmath{v} = 0,\label{eq:mass}\\
        \rho \left[\frac{\partial \bmath{v}}{\partial t} + \left(\bmath{v}\cdot \nabla \right)\bmath{v}\right] + \nabla P = 0,\label{eq:momentum}\\
        \rho T \left(\frac{\partial s}{\partial t} + \bmath{v}\nabla s \right) = -\mathcal{L},\label{eq:energy_cooling}
    \end{gather}
\end{subequations}
where $s=c_\text{P}\ln (P\rho^{-\gamma })$ is entropy per unit mass and
$\mathcal{L}$ the cooling rate. For adiabatic runs, we set $\mathcal{L}=0$. The
equation of state of ideal gas is
\begin{equation}
    P = \frac{\rho \mathcal{R} T}{\mu },
    \label{eq:ideal_gas}
\end{equation}
where $\mathcal{R}$ is the gas constant and $\mu$ the mean molecular weight.

We set up our simulations in the way similar to that in \citet{Ji2019}, so we
only briefly mention here. For our fiducial runs, we initiate a 3D domain with
$100 $\,pc in width and $300 $\,pc in height, with a resolution of $128\times
364\times 128$. Our coordinate system is set as follows: $x$ is the direction of
velocity shears, $y$ axis normal to the cold-hot interface, and $z$ the third
dimension. Outflow boundary conditions are applied to top and bottom boundaries,
and periodic boundary conditions to the rest. Cold gas and hot gas with a
velocity shear are filled in the top ($y>0$) and bottom ($y<0$) regions of the
domain respectively, with the top (cold) gas initially stationary and the bottom
(hot) gas moving at a velocity of $v_\text{hot}$. The fluid quantities of these
two phases are connected smoothly with a tanh function:
\begin{align}
    \rho (y) &= \rho_\text{hot} + \frac{\rho_\text{cold} - \rho_\text{hot} }{2} \left[1+\tanh \left(\frac{y}{a} \right) \right], \\
    v_x (y) &= \frac{v_\text{hot }}{2} \left[1-\tanh \left(\frac{y}{a} \right)\right],
\end{align}
where $v_x$ is the gas velocity along $x$ direction, and $a=2.5$\,pc the
half-length of the shear layer. This sets up a two-phase medium with cold-dense
gas on the top and hot-diffuse gas on the bottom. To trigger the Kelvin-Helmholtz
instability at the interface between the two phases, we initialize a velocity
perturbation along $y$ direction:
\begin{equation}
    v_y = v_\mathrm{pert} \exp\left[-\left(\frac{y}{a}\right)^2\right]\sin \left(\frac{2\pi x}{\lambda }\right) \sin \left(\frac{2\pi z}{\lambda }\right),
    \label{eq:init_perturb}
\end{equation}
where $\lambda $ is the perturbation wavelength, and $v_\mathrm{pert}$ is set to
$v_\text{hot}/100$. The parameters for the fiducial run are listed in
Table~\ref{tab:param_fiducial}. With the parameters for conditions, the cooling
time $t_\mathrm{cool,mix}=70\,$Myr for $T_\mathrm{mix}=2\times 10^{5}\,$K in our
setup.\footnote{By following the choice of $T_\text{mix}$ (thus
$t_\mathrm{cool,mix}$) in \citet{Tan2021}, we can calculate the Damk{\"o}hler
numbers in our simulations (see \S\ref{sec:caveats}). Though
$t_\mathrm{cool,mix}$  defined here is relatively large, the evolution times we
use ($\ge 100\,$Myr) have turned out to be sufficient for a saturated state of
the mixing layer to arise. In case the cooling time for the geometric mean of
hot and cold temperatures is of interest, $t_\text{cool} = 10\,$Myr for
$T=10^5\,$K.}

\begin{table*}
    \centering
    \caption{Parameters for the fiducial hydrodynamic simulation. The speed of
        sound in the hot medium $c_\text{s} = 117.72$\,km/s. With $\delta =
        100$, the critical Mach number for the KH instability suppression
        $\mathcal{M}_\text{crit} = 1.34$.}
    \label{tab:param_fiducial}
    \begin{tabular}{c c c c c c c}
        \toprule[1pt]\midrule[0.4pt] {\bf Domain size} & {\bf Resolution} & {\bf
             Metallicity}
             & {\bf Number density} &
             {\bf Velocity shear} &
             {\bf Temperature}
             & {\bf Perturbation
             wavelength}
             \\
        (pc)       &
        & ($Z_{\sun}$)         & ($\mathrm{cm}^{-3}$) &
        ($\mathrm{km}/\mathrm{s}$)     & ($\mathrm{K}$)
        & ($\mathrm{pc}$)          \\
        \midrule[0.4pt] $100\times 300\times 100$      & $128\times 384\times
        128$       & 0.1
        &
        \begin{tabular}{c}
      $n_\text{cold}=10^{-2}$ \\ $n_\text{hot}=10^{-4}$\end{tabular} &
        \begin{tabular}{c}
      $100$ \\
      ($\mathcal{M} = 0.85$)
        \end{tabular}
             &
        \begin{tabular}{c}
      $T_\text{cold} = 10^4$ \\
      $T_\text{hot} = 10^6$\end{tabular}     & $\lambda =
        100$ \\
        \bottomrule[1pt]
    \end{tabular}
\end{table*}

For the simulations with radiative cooling included, we use a set of cooling
functions which assumes collisional and photoionization equilibrium, instead of
summing up losses from all the time-dependent ion abundances, since the latter
requires excessive memory and computational power. The cooling curves with
photoionization are generated by the spectral synthesis code {\small CLOUDY}
with background radiation intensity from Madau \& Haardt 2005 at the redshift of
0. For more details about the techniques, see \citet[Figure 1]{Ji2019}.

To study how the Mach number of shear velocity affect the evolution of turbulent
mixing layers, we have run a series of simulations with different initial hot
gas velocities, and parameters for which are listed in
Table~\ref{tab:param_shear}. Since the TML thickness can expand significantly
with greater Mach numbers, we adopt varying domain heights to ensure the entire
TMLs are completely enclosed by the simulation domains. In addition, we carry
out adiabatic counterparts for high-Mach number runs to probe the possible
impact of cooling on the early growth of turbulence, and a suite of low- and
high-resolution simulations for convergence studies, and a few runs with wider
simulation boxes to check the effects of box width ($x$ direction), with
parameters listed in Table~\ref{tab:special}. Finally, Table~\ref{tab:notation}
summarizes the notations used in the paper for clarity purposes.

\begin{table}
    \centering
    \caption{Parameters for simulations with different initial Mach numbers. Other parameters (not listed here) for each run are identical to that for the fiducial case.}
    \label{tab:param_shear}
    \begin{tabular}{c c c c }
        \toprule[1pt]\midrule[0.4pt] {\bf Velocity shear} & {\bf Mach number} &
        {\bf Domain height}     & {\bf Name} \\
        (km/s)        & $\mathcal{M}$     &
        (pc) & \\ \midrule[0.4pt] 10         & 0.09      &
        300  & \tt M0.09
        \\
        20         & 0.17      &
        300  & {\tt M0.17}
        \\
        50            & 0.42    &
        300  & \tt M0.42 \\
        100           & 0.85    &
      300  & {\tt FC} or {\tt M0.85}
      \\
        200           & 1.70    &
        400  & \tt M1.70 \\
        300           & 2.55    &
        600  & \tt M2.55 \\
        500           & 4.25    &
        800  & \tt M4.25 \\
        800           & 6.80    &
        900  & \tt M6.80 \\
        \bottomrule[1pt]
    \end{tabular}
    
\end{table}

\begin{table}
    \centering
    \caption{Parameters for studies of the impact of cooling and resolution. The
    adiabatic runs are marked with ``{\tt NC}'' (no cooling). Other parameters
    (not listed here) for each run are identical to that for its counterpart in
    Table~\ref{tab:param_shear}. We use $\Delta_\text{FC}/\Delta $ to represent
    resolution here (the grid scale of the fiducial run $\Delta_\text{FC} =
    0.78\,$pc)}
    \label{tab:special}
    \begin{tabular}{c c c}
        \toprule[1pt]\midrule[0.4pt] {\bf Velocity shear} & {\bf Feature} & {\bf Name}
        \\
        (km/s)     &  &       \\ \midrule[0.4pt] 100
        & No radiative cooling    & {\tt M0.85\_NC}
        \\
        200        &  No radiative cooling    
        & \tt M1.70\_NC   \\
        300        &  No radiative cooling    
        & \tt M2.55\_NC   \\
        500        &  No radiative cooling    
        & \tt M4.25\_NC   \\
        800        &  No radiative cooling    
        & \tt M6.80\_NC   \\
        100        & Resolution: 0.5
        & {\tt M0.85\_LR} \\
        200        & Resolution: 0.5
        & \tt M1.70\_LR   \\
        200        & Resolution: 2.0
        & \tt M1.70\_HR   \\
        300        & Resolution: 0.5
        & \tt M2.55\_LR   \\
        300        & Resolution: 2.0
        & \tt M2.55\_HR   \\
        200         & {Box width: 200 pc} 
        & {\tt M1.70\_W2} \\
        200        & {Box width: 300 pc} 
        & {\tt M1.70\_W3}  \\
        300        & {Box width: 300 pc} 
        & {\tt M2.55\_W3}  \\
        \bottomrule[1pt]
    \end{tabular}
    
\end{table}

\begin{table*}
    \centering
    \caption{Notation system of energy fluxes/rates discussed in this paper.}
    \label{tab:notation}
    \begin{tabular}{c c c}
        \toprule[1pt]\midrule[0.4pt] {\bf Symbol} & {\bf Description/Definition}
        & {\bf Note}          \\
        \midrule[0.4pt] $\mathcal{L}$         &
        \begin{tabular}{p{6cm}} Radiative cooling rate; $\mathcal{L}=
        n^2\Lambda(n,T)$. \end{tabular} & \begin{tabular}{p{8cm}} The energy
        loss rate via radiative cooling (emissivities) \end{tabular} \\
        \midrule[0.4pt] $\mathcal{H}$         &
        \begin{tabular}{p{6cm}} Enthalpy loss rate; $\mathcal{H}=
        -\bar{P}\tilde{v}_{i,i}-\left< Pv_{i,i}' \right>$. \end{tabular} &
        \begin{tabular}{p{8cm}} The rate at which gas enthalpy is being
        converted to other forms; the angle brackets denote the ensemble
        averages at the same $y$ coordinate (in our simulations); negative
        $\mathcal{H}$ indicates gas expansion, rather than
        compression.\end{tabular} \\  \midrule[0.4pt] $\mathcal{D}$         &
        \begin{tabular}{p{6cm}} Turbulent dissipation rate; $\mathcal{D}\equiv
        -t_{ij}^\text{R}\tilde{S}_{ij}$, where $t_{ij}^\text{R}=-\left< \rho
        v_i' v_j'\right>$ and $\tilde{S}_{ij} = \frac{1}{2}\left(
        \tilde{v}_{i,j} + \tilde{v}_{j,i} -
        \frac{2}{3}\delta_{ij}\tilde{v}_{k,k} \right)$. \end{tabular} &
        \begin{tabular}{p{8cm}} $v_i = \tilde{v}_i + v_i'$, $\left<\rho
        v_i'\right>=0$, where $\tilde{v}_i$ and $v_i'$ are mean and fluctuating
        components respectively; $t_{ij}^\text{R}$ is the Reynolds stress, and
        $\tilde{v}_{i,j}$ and $\tilde{v}_{j,i}$ denote partial derivatives
        ($\partial_j$ and $\partial _i$) of $\tilde{v}_i$ and $\tilde{v}_j$
        respectively. \end{tabular} \\   \midrule[0.4pt] $Q$ &
        \begin{tabular}{p{6cm}}Surface brightness of radiative mixing layers;
        $Q\equiv \int_{y_1}^{y_2} n^2 \Lambda \text{d}y $. \end{tabular} &
        \begin{tabular}{p{8cm}}The cooling region is included in the range from
        $y_1$ to $y_2$.\end{tabular} \\  \midrule[0.4pt] $F_\text{h}$
        &
        \begin{tabular}{p{6cm}} Enthalpy flux; $F_\text{h}\equiv
        5/2Pv_\text{in}$.
        \end{tabular}
        & \begin{tabular}{p{8cm}} $v_\text{in}$ is the inflow velocity of hot
        gas (with respect to the TML). In our calculation, $v_\mathrm{in} =
        v_{y,\mathrm{bot}} - v_{y,\mathrm{inter}}$, i.e., the inflow velocity is
        obtained by measuring the gas $y$-velocity across the bottom boundary of
        the box $v_{y,\mathrm{bot}}$, relatively to the migrating velocity of
        the TML interface $v_{y,\mathrm{inter}}$. \end{tabular}    \\
        \midrule[0.4pt] $F_\text{k}$    &
        \begin{tabular}{p{6cm}} Flux of bulk kinetic energy; $F_\text{k}\equiv
      \epsilon_\text{k}v_\text{in} = \frac{1}{2}\rho \bmath{v}^2
      v_\text{in}$.\end{tabular}      &
        \begin{tabular}{p{8cm}} $\epsilon_\text{k}$ is kinetic energy density
      (erg\,cm$^{-3}$); specially, the flux into the TML,
      $F_\text{k}\sim 5/2Pv_\text{in}\times
      1/3\mathcal{M}^2$.\end{tabular} \\  \midrule[0.4pt] $F_\text{h+k}$
      &
        \begin{tabular}{p{6cm}} $F_\text{h+k} \equiv F_\text{h} + F_\text{k}$
        \end{tabular}
        &
        \begin{tabular}{p{8cm}} When TMLs reach equilibrium,
      $F_\text{h+k}\sim Q$ (i.e., Eq.~\eqref{eq:balance}).
      \end{tabular}
      \\  \midrule[0.4pt] $H$             &
        \begin{tabular}{p{6cm}} Enthalpy loss rate integrated along $y$
        direction; $H\equiv \int_{y_1}^{y_2}\mathcal{H} \text{d}y$.
        \end{tabular}
        &
        \begin{tabular}{p{8cm}} The rate at which enthalpy is consumed through
        the compression of gas; [$y_1,y_2$] accommodates the entire turbulent
        region, and we choose the bottom and top boundaries as $y_1$ and $y_2$ in our calculation in practice. Note that $H
        \sim F_\mathrm{h}$ only holds when TMLs reach quasi-equilibrium
        (Eq.~\eqref{eq:fine_balance}), while it is not always the case -- see
        \S\ref{sec:energies} for detailed discussions. \end{tabular} \\
        \midrule[0.4pt] $D$             &
        \begin{tabular}{p{6cm}} Turbulent dissipation integrated along $y$
        direction; $D\equiv \int_{y_1}^{y_2}\mathcal{D} \text{d}y$.
        \end{tabular}      &
        \begin{tabular}{p{8cm}} The turbulent region is contained in the range
      from $y_1$ to $y_2$; $D$ equals to the loss rate of kinetic energy in the
      whole turbulent region with a unit of erg\,s$^{-1}$\,cm$^{-2}$ (can be
      directly compared with energy fluxes above). Note that since to
      accurately calculate turbulent dissipation from the definition is tricky,
      here we instead derive $D$ from the conservation of energy, i.e., to
      subtract the (kinetic + thermal) energy change due to radiative cooling
      and inflow/outflow across domain boundaries, from the total energy change
      integrated over the whole simulation box.
    \end{tabular}

        \\
        \bottomrule[1pt]
    \end{tabular}
\end{table*}

\section{Results}
\label{sec:results}

\subsection{Short-time evolution: suppression and revival of KH instability}
\label{sec:time_suppress}

\begin{figure*}
    \includegraphics[width=0.503\textwidth, trim= 0 0 30 0,
        clip]{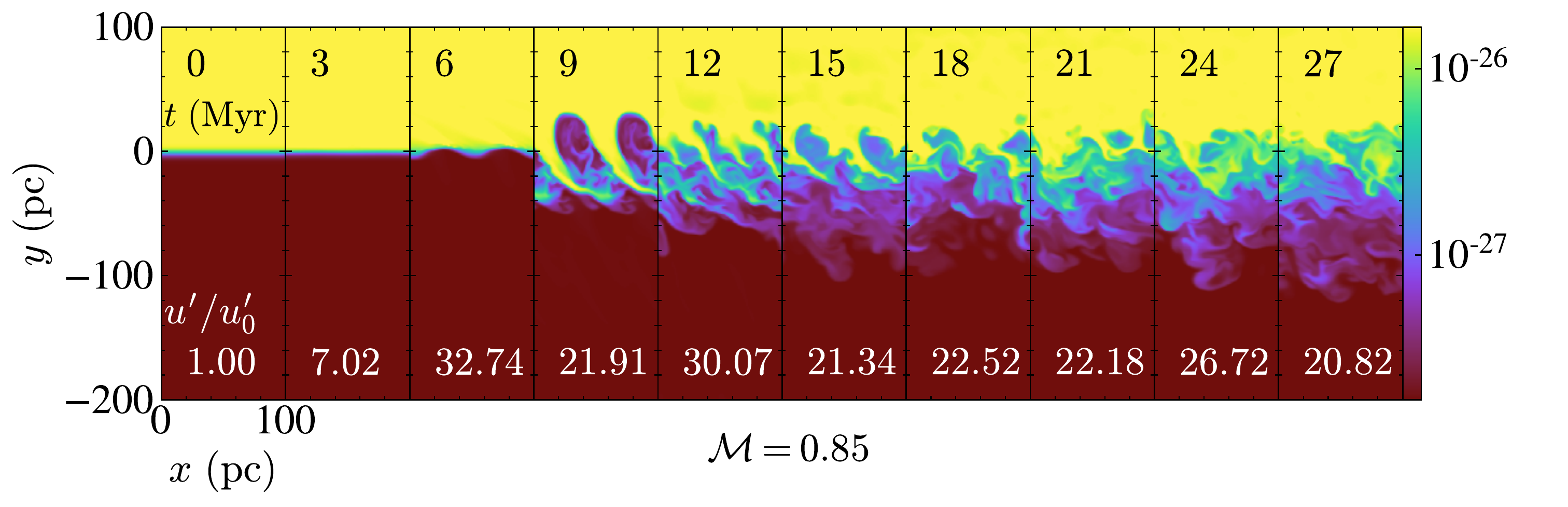}%
    \includegraphics[width=0.497\textwidth]{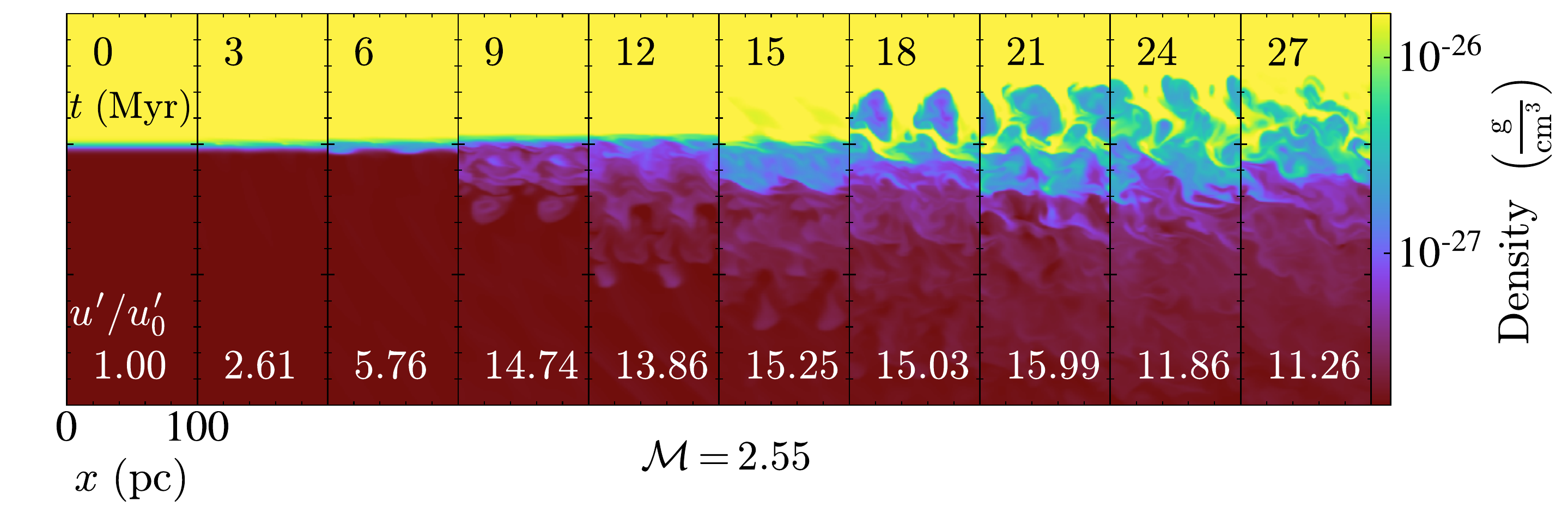}
    \caption{Slice plots of density evolution in {\tt M0.85} (left) and {\tt
      M2.55} (right) during the early stage of evolution ($t\leq
      27\,\mathrm{Myr}$), labeled with current evolution time and the
      turbulent velocity $u'$ normalized to the magnitude of initial
      velocity perturbation $u_0'$. Compared to the low-Mach number
      case {\tt M0.85} where the KH instability is already
      well-developed at $t\sim 9\,\mathrm{Myr}$, the KH instability of long-wavelength perturbations in
      the high-Mach number case {\tt M2.55} is clearly suppressed
      until $t\sim 18\,\mathrm{Myr}$ while small-scale turbulence begins to
      develop earlier.}
    \label{fig:slice_early_evo}
\end{figure*}

Fig.~\ref{fig:slice_early_evo} shows the time evolution of gas density in two
representative simulations: the low-Mach number case {\tt M0.85} (left) and the
high-Mach number case {\tt M2.55} (right). For {\tt M0.85}, it is just the
``standard' version of the KH instability with radiative cooling: initial
perturbations with long wavelengths grow significantly until secondary
instabilities start to appear at $t\sim 12\,\mathrm{Myr}$, which results in
long-wavelength structure cascading to smaller scales. The turbulent mixing
layer ultimately saturates with an approximately fixed thickness after $t\sim
21\,\mathrm{Myr}$, when the radiative cooling balances the hot gas entrainment
\citep{Ji2019}. However, the high-Mach number case of {\tt M2.55} is different
in two aspects. First, since there are no unstable modes at
$\mathcal{M}>\mathcal{M}_\text{crit}$ \citep{Mandelker2016}, large eddies do not
appear at the cold-hot gas interface initially. At $t\sim 9\,$Myr, it is from
small scales that the TML begins to grow, since the initial long-wavelength
perturbations are stabilized by high velocity shears and already damped. Second,
in contrast to a fixed thickness in {\tt M0.85}, the high-Mach number turbulent
mixing layer ({\tt M2.55}) continues growing in thickness and does not reach a
saturation state at least up to $t \sim 27\,\mathrm{Myr}$.

\begin{figure}
    \includegraphics[width=\columnwidth]{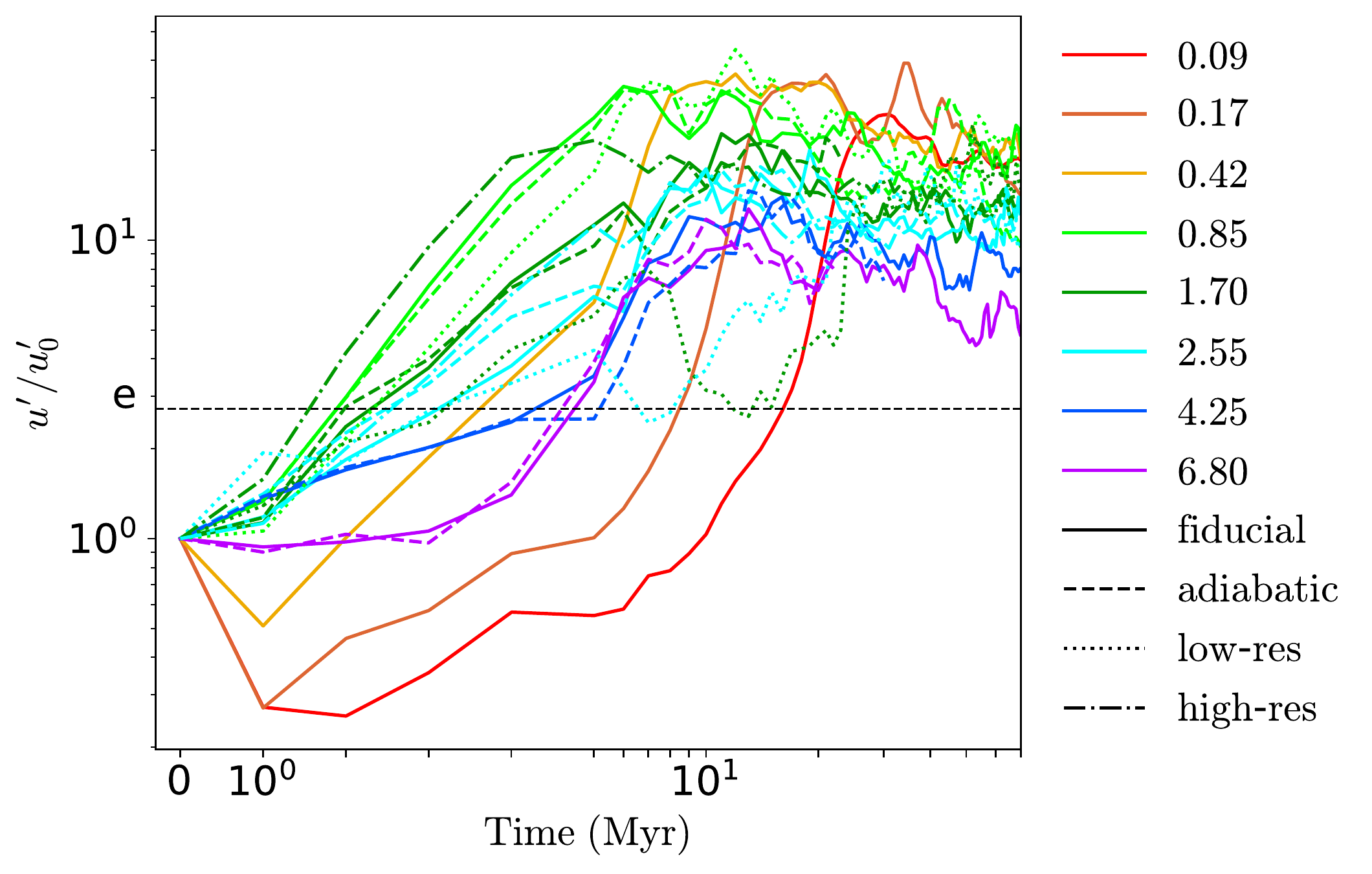}
    \caption{Time evolution of peak turbulent velocity (normalized to the
    magnitude of initial perturbations) in runs with different Mach numbers
    ranging from $0.09$ to $6.8$, represented by different colors. The solid
    lines refer to runs with radiative cooling included, and the dashed ones are
    their adiabatic counterparts (i.e., without cooling). The dotted lines are
    low-resolution runs: {\tt M0.85\_LR} (lime), {\tt M1.70\_LR} (green) and
    {\tt M2.55\_LR} (aqua), and the two dash-dotted lines are high-resolution
    runs: {\tt M1.70\_HR} (green) and {\tt M2.55\_HR} (aqua). A factor of $e$ is
    annotated on the $y$-axis, indicating the e-folding time of the peak
    turbulent velocity in each run. As expected, for the three low-Mach number
    runs, the growth rate increases with raising $\mathcal{M}$ (shorter $t_e$).
    In contrast, in high-Mach number cases, runs with larger $\mathcal{M}$ have
    longer e-folding time, and the effect of radiative cooling on e-folding time
    is negligible. The e-folding time of turbulence growth is
    resolution-independent (resolution-dependent) at low (high) Mach numbers.}
    \label{fig:vt_evo}
\end{figure}

We investigate the KH instability quantitatively during the initial suppression
phase in Fig.~\ref{fig:vt_evo}, where the peak turbulent
velocity\footnote{In practice, following \citet{Ji2019}, we use the $v_y$ velocity
dispersion $\delta v_y \equiv \sqrt{\langle v_y^2 \rangle - \langle v_y
\rangle^2}$ to measure the
turbulent velocity, in order to exclude the contribution of the bulk shear
velocity predominantly along the $x$-direction.} (normalized to
the initial velocity perturbation) $u'/u'_0$ is plotted as a function of
simulation time. Different colors represent runs with different Mach numbers
ranging from $0.09$ to $6.8$, and the solid (dashed) lines are runs with
(without) radiative cooling. For low-Mach number cases ($\mathcal{M}<0.9$), as
expected, the growth rates of turbulence increase with greater Mach number, due
to faster growth rate of the KH instability. In contrast, at high Mach numbers,
the growth rates become inversely-correlated with Mach numbers, indicating
stronger suppression of turbulence at higher Mach numbers when
$\mathcal{M}\gtrsim \mathcal{M}_\mathrm{crit}$. However, due to the
existence of numerical (in our simulations) or physical (in reality) viscosity,
the sharp discontinuity in both densities and velocities smooth out, bringing
local shears below the critical Mach number, and thus the KH instability revives
and triggers turbulence.

Besides, since the initial long-wavelength perturbations have already been
damped in the high-Mach number case, the KH instability has to grow from
smallest scales (where the linear growth rate is maximized) which are determined
by the numerical resolution. Therefore, the e-folding time of turbulence growth
in high-Mach number cases is unfortunately resolution-dependent:
higher-resolution runs where smaller scales are resolved have faster turbulence
growth and thus shorter e-folding time compared with their low-resolution
counterparts (e.g., see {\tt M1.70\_LR}, {\tt M1.70} and {\tt M1.70\_HR} in
Fig.~\ref{fig:vt_evo}). However, since the KH instability in low-Mach number
cases grows from unsuppressed long-wavelength perturbations, simulations can
converge much more easily as long as initial perturbations are well-resolved.
Indeed, Fig.~\ref{fig:vt_evo} shows that the e-folding time in {\tt M0.85} and
{\tt M0.85\_LR} is comparable, which is expected by the analysis above. We also
find that cooling has little influence on the initial stage of turbulence
growth, which is consistent with the findings in \citet{Ji2019} that cooling
only becomes important in the saturation stage, but not for the ``kick-off'' of
turbulent mixing layers (which is triggered by the KH instability).

We note that although, as discussed above, the e-folding time of turbulent
velocity growth in the high-Mach number case does not converge, TML properties
in final saturation stages do converge, which is shown by convergence tests in
\S\ref{sec:convergence}. For our following discussion, we stick to TML
properties in the saturation stage.

\subsection{Long-term evolution: morphologies and surface brightness}

\begin{figure*}
    \centering
    \includegraphics[width=1.0\textwidth]{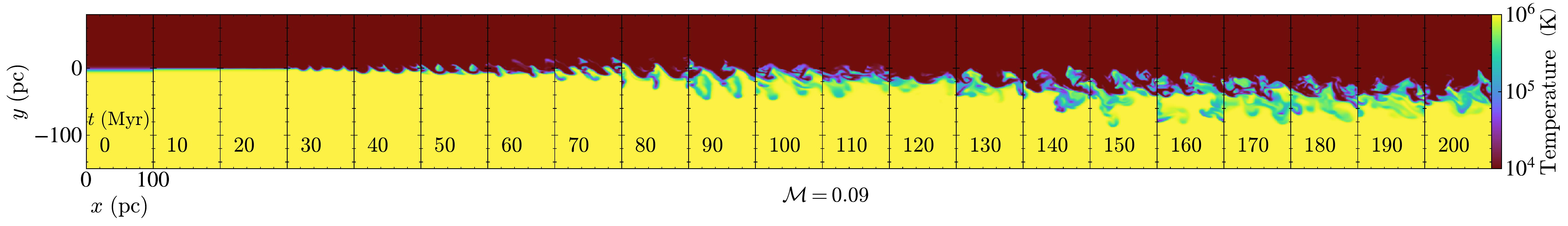}\\
    \includegraphics[width=1.0\textwidth]{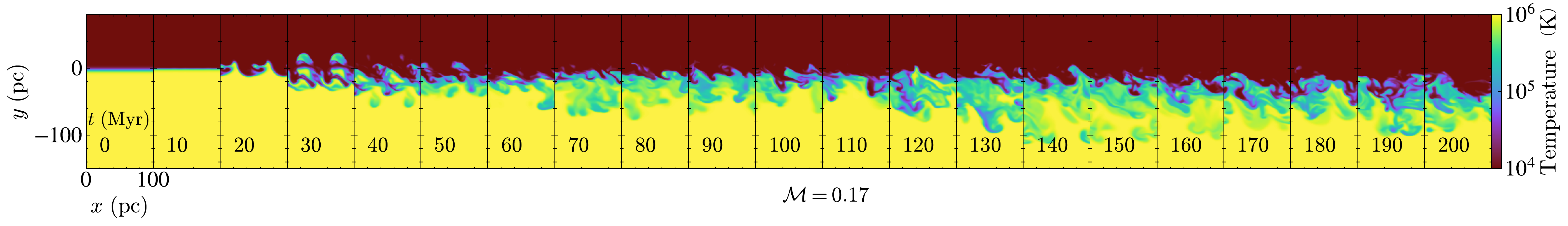}\\
    \includegraphics[width=0.508\textwidth, trim= 0 0 30 0,
        clip]{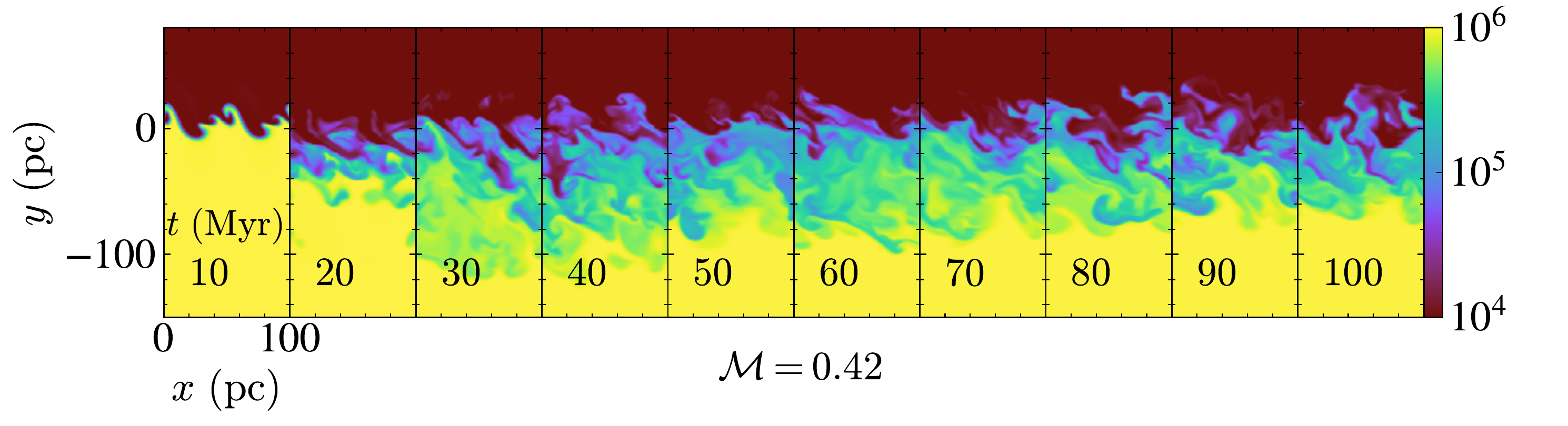}%
    \includegraphics[width=0.492\textwidth]{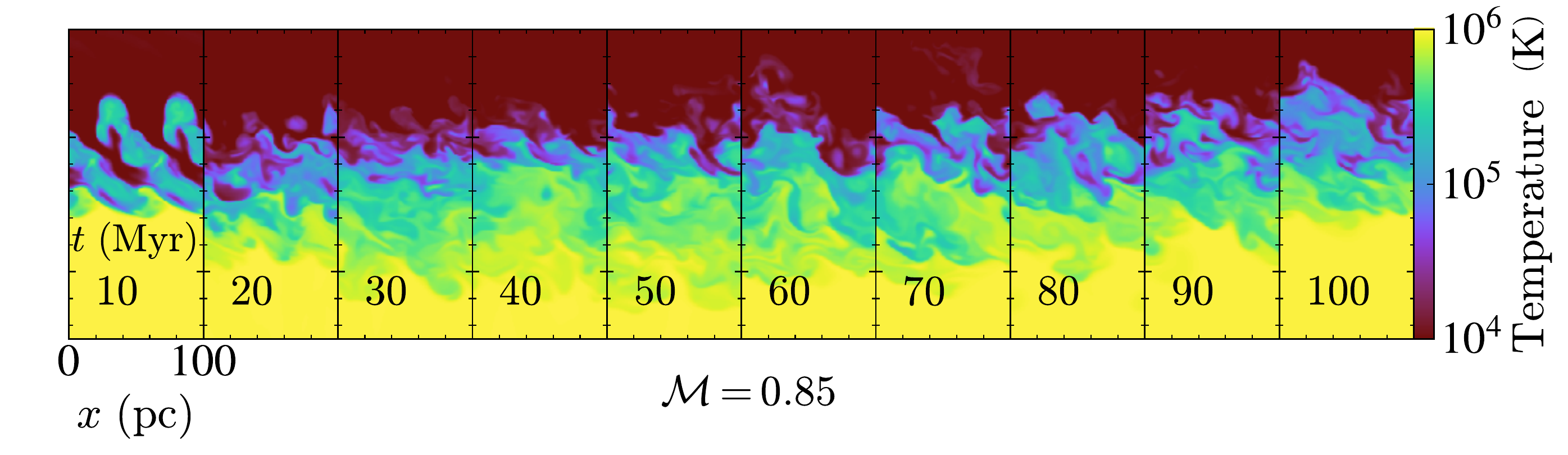}\\
    \includegraphics[width=0.508\textwidth, trim= 0 0 30 0,
        clip]{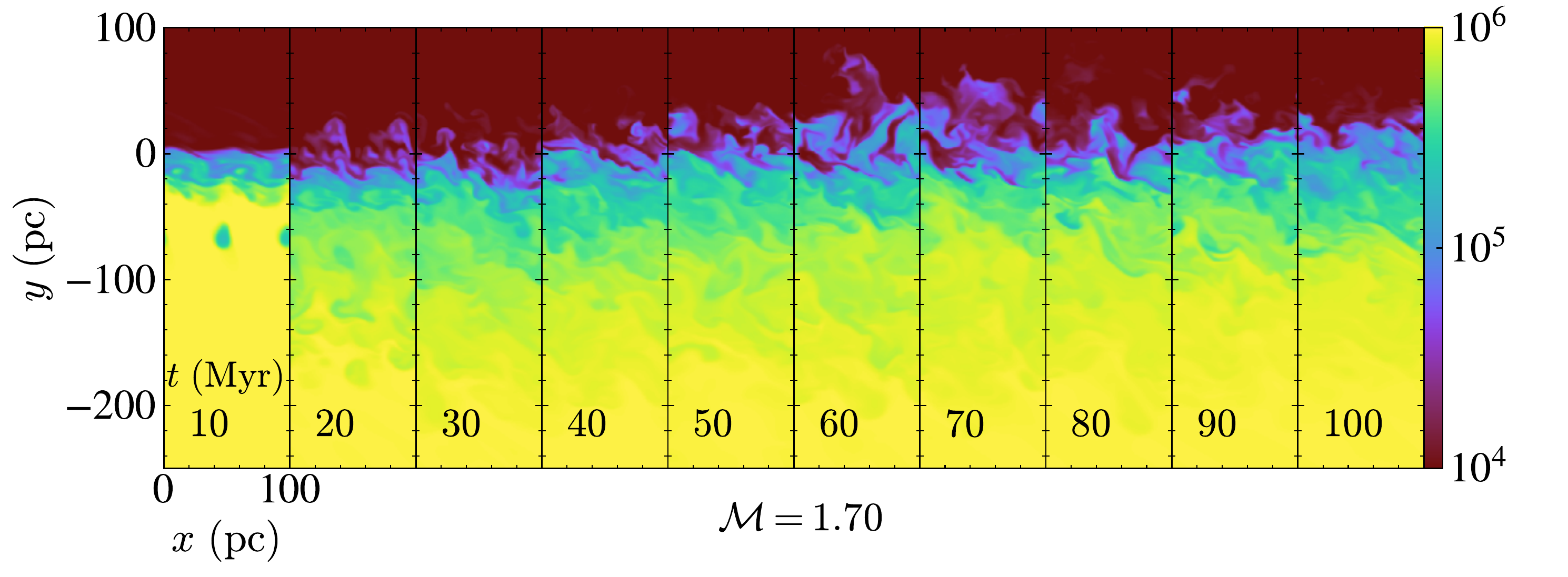}%
    \includegraphics[width=0.492\textwidth]{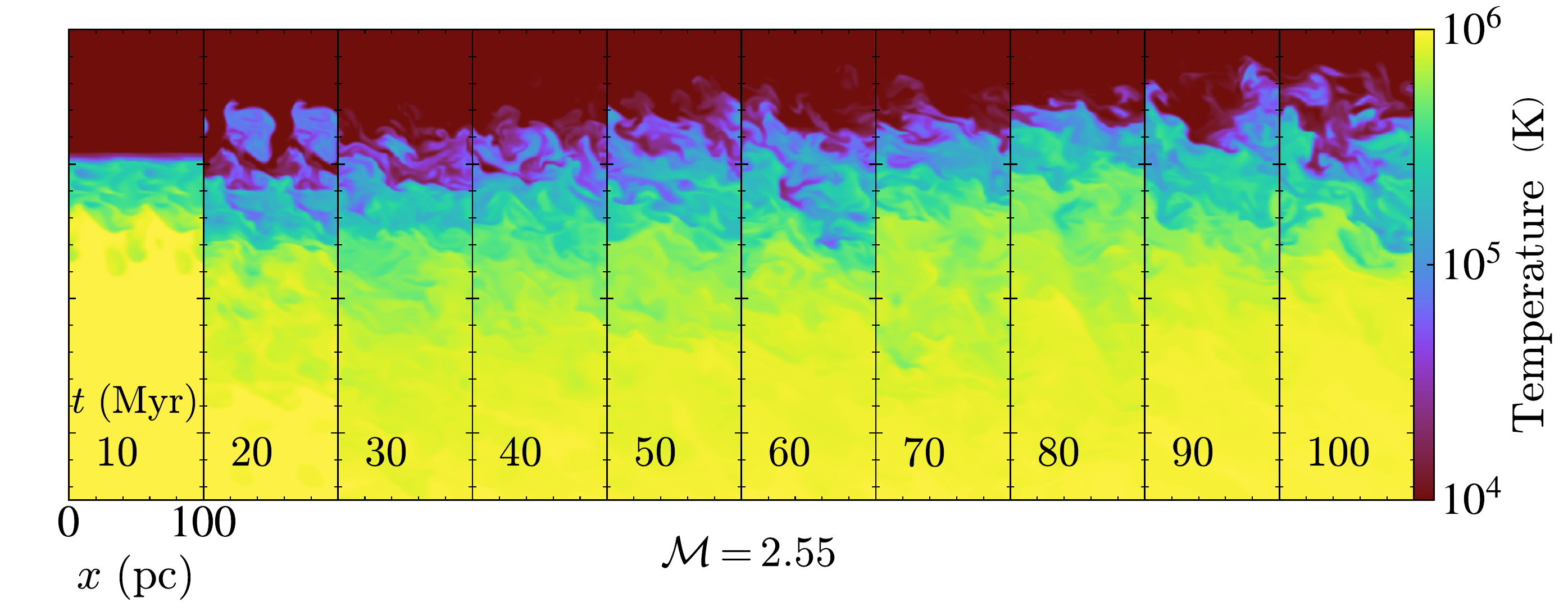}\\
    \includegraphics[width=0.508\textwidth, trim= 0 0 30 0,
        clip]{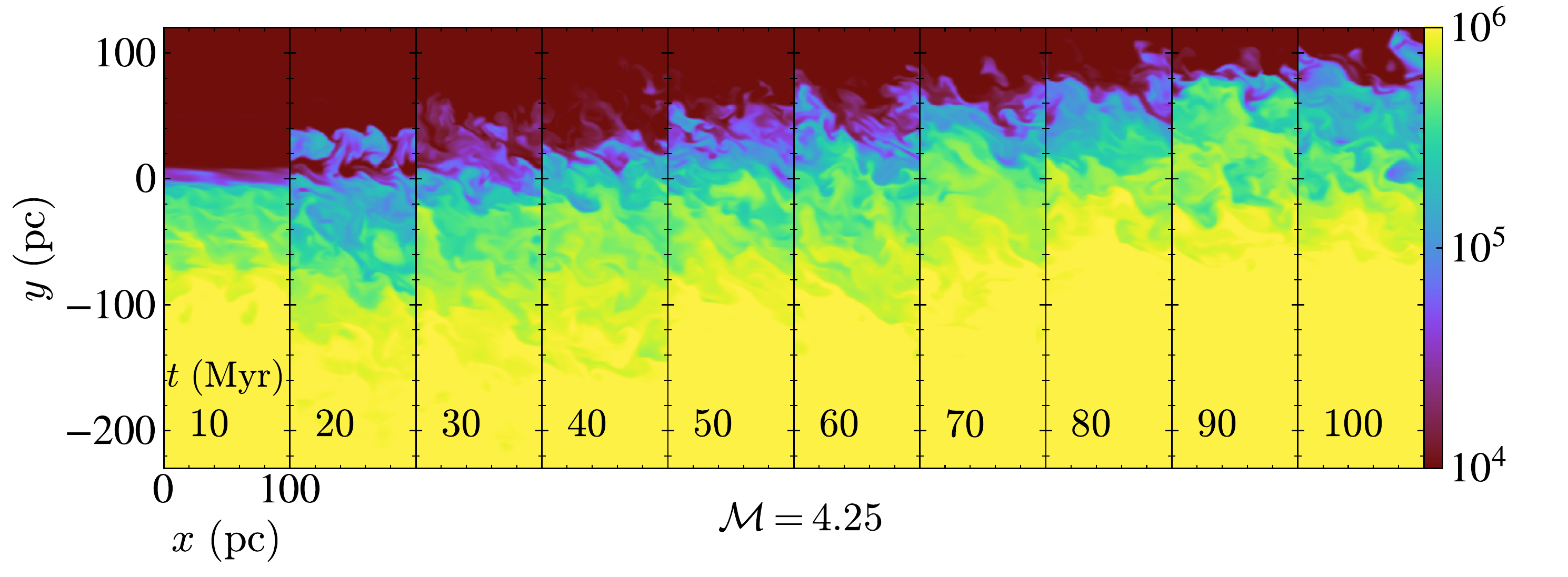}%
    \includegraphics[width=0.492\textwidth]{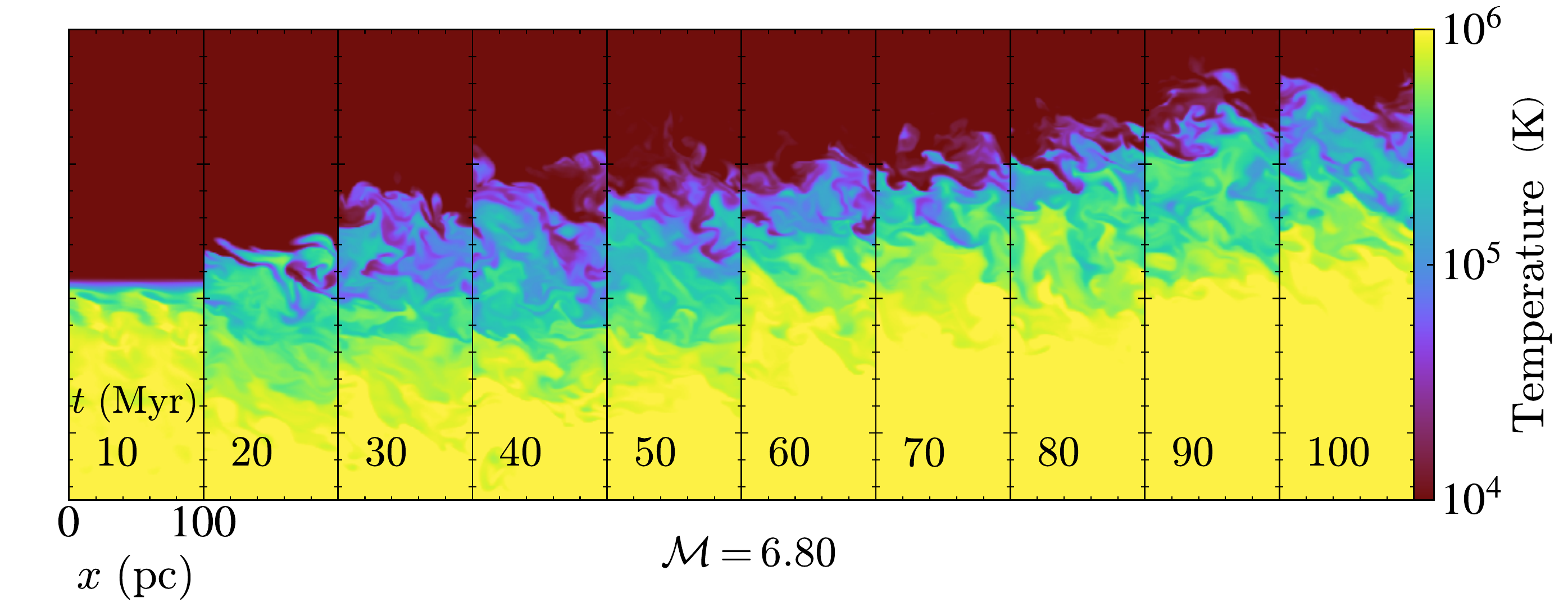}
    \caption{Slice plots of temperature evolution in runs {\tt M0.09}, {\tt M0.17,} {\tt
      M0.42}, {\tt M0.85}, {\tt M1.70}, {\tt M2.55}, {\tt M4.25} and
      {\tt M6.80}. For subsonic runs, the saturated mixing layer
      thickness increases with Mach number, while for supersonic ones,
      the thickness remains roughly constant. \textit{Note:} (i) The
      slices have been cropped accordingly to save space, so the boxes
      appear shorter than their authentic sizes (regions with
      little temperature variation are cut off). (ii) Though the
      intermediate temperature volumes in {\tt M1.70} and {\tt M2.55}
      are seemingly more extensive than those in {\tt M4.25} and {\tt
      M6.80}, their cooling zone widths are actually approximately
      equal, for the reason that $T\gtrsim 3\times 10^5\,$K gas nearly
      does not cool itself. Meanwhile, the temperature of part of the
      turbulent zones in {\tt M4.25} and {\tt M6.80} goes beyond
      $10^6\,$K owing to turbulent dissipation.}
    \label{fig:temp_slices}
\end{figure*}

Fig.~\ref{fig:temp_slices} shows the time evolution of temperature slices in
simulations with Mach numbers ranging from $0.09$ up to $6.80$. In the run with
smallest velocity shear {\tt M0.09}, the TML barely grows until $t\gtrsim
140\,\mathrm{Myr}$ with apparently much thinner thickness of $\sim
20\,\mathrm{pc}$. With larger velocity shears, the TMLs in {\tt M0.42} and {\tt M0.85} already develop at $t\sim 30\,\mathrm{Myr}$ with larger thickness of $\sim
100\,\mathrm{pc}$. For supersonic runs of {\tt M1.70}, {\tt M2.55}, {\tt M4.25}
and {\tt M6.80}, after the initial development of turbulence, we do not
find the saturated TML thickness grows significantly with increasing Mach
number, e.g., the run {\tt M6.80} has a velocity shear of $\sim 4$ times larger
than {\tt M1.70}, while ultimately only reaches saturated TML thickness of $\sim
100\,\mathrm{pc}$ which is comparable with {\tt M1.70}.

\begin{figure}
    \centering
    \includegraphics[width=\columnwidth]{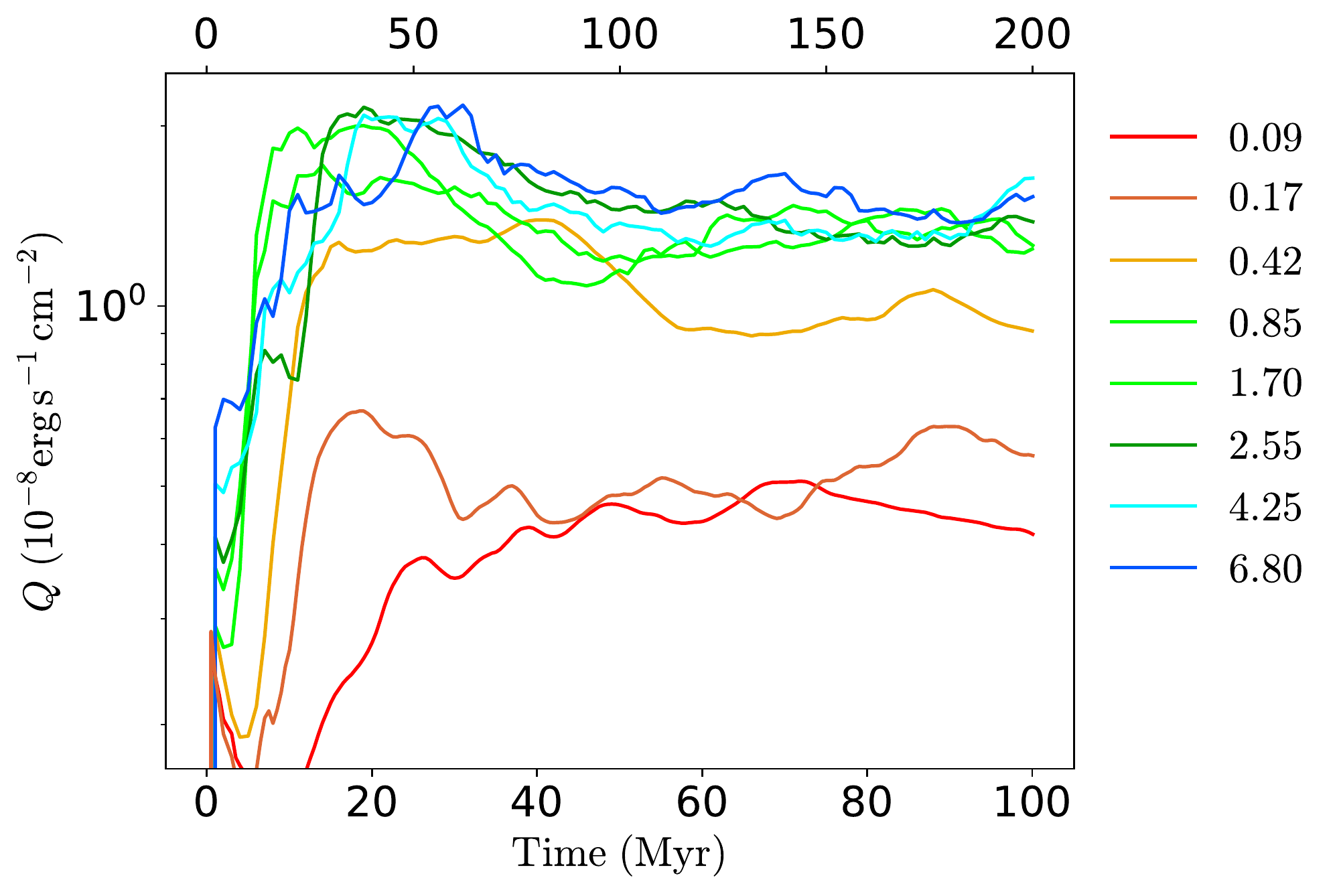}
    \caption{Time evolution of surface brightness from simulations with
        different Mach numbers (represented by colors). For {\tt M0.09} and {\tt
        M0.17}, a wider time axis (ranging up to $200\,\mathrm{Myr}$) located at
        the top of the figure is used, since they evolve much slower. All runs
        reach a steady state with a saturated surface brightness in the final
        stage. Saturated surface brightness increases with larger Mach numbers
        at $\mathcal{M}\lesssim 1$, while ceases to increase further with Mach
        numbers at $\mathcal{M} \gtrsim 1$.}
    \label{fig:bright_evo}
\end{figure}

Quantitatively, Fig. \ref{fig:bright_evo} shows the time evolution of surface
brightness $Q$ in simulations with different Mach numbers. In all runs, the
saturated surface brightness finally saturates after an initial growth stage
ranging from $\sim 30$ to $100\,\mathrm{Myr}$ (depending on initial Mach
numbers), suggesting a steady state is indeed achieved. For subsonic runs {\tt
M0.09}, {\tt M0.17}, {\tt M0.42} and {\tt M0.85}, the final saturated surface brightness is
positively correlated with Mach number, but for supersonic ones, the saturated
surface brightness remains at a similar level, regardless of varying Mach
numbers ranging from $1.7$ up to $6.8$. This is consistent with the
morphological evolution where the thickness of TMLs ceases increasing with
raising Mach number. We will discuss its cause and consequences in the following
sections.

\subsection{Velocity profiles tracing sound speed}
\label{sec:velocity_constraint}

\begin{figure}
    \centering
    \includegraphics[width=\columnwidth]{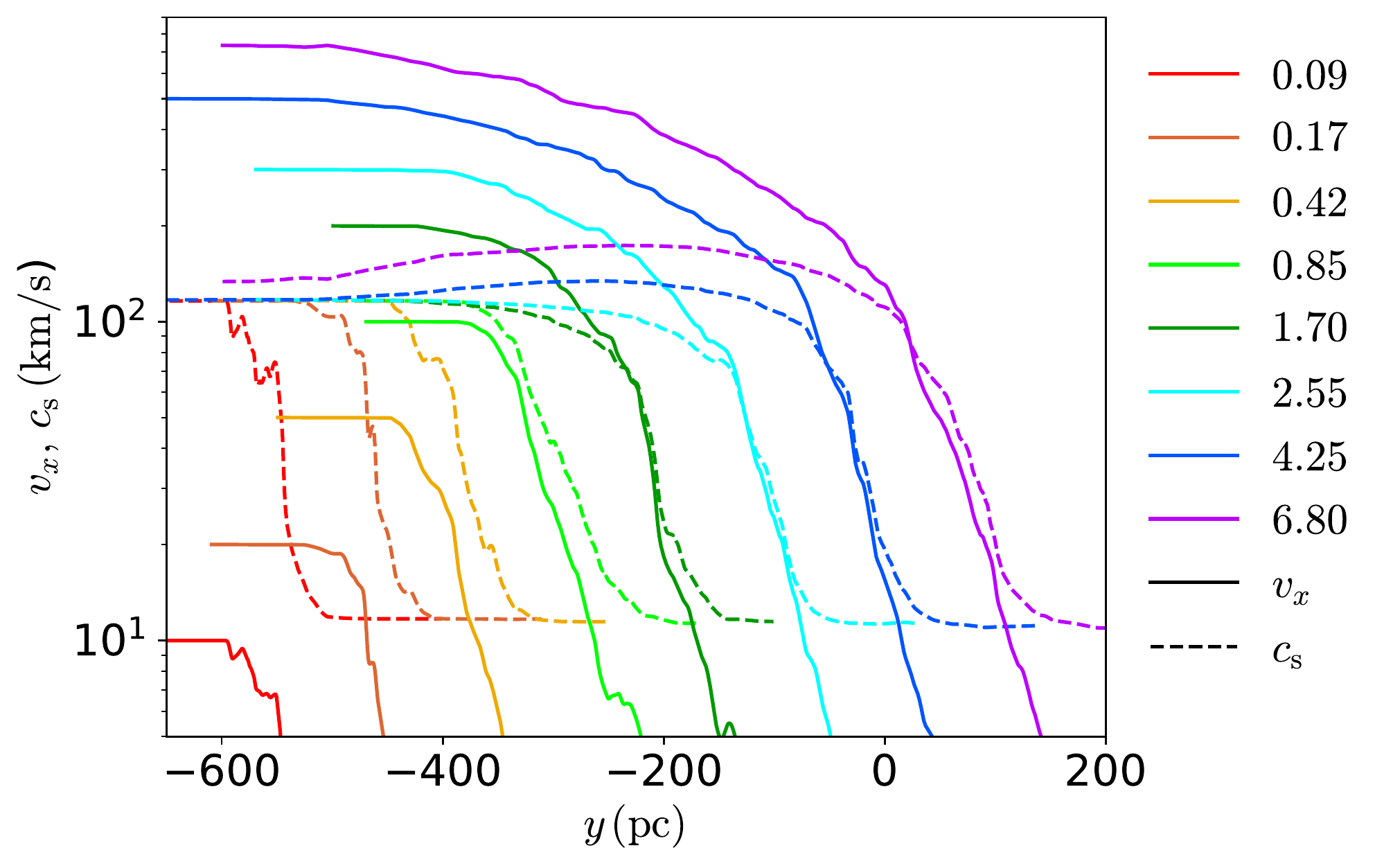}
    \caption{Profiles of $x$-velocity (solid) and sound speed $c_\text{s}$ (dashed) in
        the saturation stage ($t=80\,$Myr, except $t=160\,$Myr is chosen for the
        runs {\tt M0.09} and {\tt M0.17} due to their relatively slow turbulence growth). Apparently in each
        of these runs, $v_x$ does not exceed local sound speed in the mixing region of the TML, which are traced by intermediate sound speed. Particularly, $v_x$
        curves trace corresponding $c_\text{s}$ curves in supersonic cases.
        \textit{Note:} For clearness, the positions of curve pairs ($v_x(y)$ and
        $c_\text{s}(y)$) have been shifted properly along the horizontal axis here, instead of using original coordinates. }
    \label{fig:vx_compare}
\end{figure}

According to mixing length theories, turbulent velocities are proportional to
local velocity gradients ($u'\propto \partial v_x / \partial y$), which is
demonstrated to be applicable to turbulent mixing layers by \citet{Tan2021}.
Therefore, we start with inspecting velocity profiles in
Fig.~\ref{fig:vx_compare}, where volume-averaged profiles\footnote{Since
$\tau_\text{turb} < t_\text{cool}$, the TML with slow cooling (``well-stirred
reactor'') appears to be single-phase (Fig.~\ref{fig:temp_slices}, see also
Figure 7 of \citealt{Tan2021}), i.e., the temperature distribution of a certain
slice (fixed $y$) of gas is relatively uniform; therefore, volume-averaged
profiles are able to track local fluid properties fairly accurately. In
contrast, fast cooling produces multi-phase mixing layers \citep{Tan2021}, where
averaged values might not be representative.} of the bulk velocity $v_x$ (solid)
and the sound speed $c_\text{s}$ (dashed) from simulations with different Mach
numbers in the saturation stage are plotted. For subsonic runs {\tt M0.09}, {\tt
M0.42} and {\tt M0.85}, the difference between $v_x$ and $c_\text{s}$ narrows
with increasing Mach number, while for supersonic runs {\tt M1.70}, {\tt M2.55},
{\tt M4.25} and {\tt M6.80}, $v_x$ profiles cease to increase with Mach number,
after reaching the local sound speed $c_\text{s}$ in regions with efficient
mixing and cooling (which are traced by intermediate sound speed).
Apparently, in supersonic cases, profiles of $x$-velocity are limited and thus
shaped by profiles of local sound speed ($v_x \lesssim c_\text{s}$), so the
mixing (cooling) regions, as shear layers, have $v_{x,\text{max}}<c_\text{s,h}$
(see $v_x$-$c_\text{s}$ intersection points in Fig.~\ref{fig:vx_compare}, which
roughly indicate positions of boundaries of the cooling region).\footnote{More
precisely, as the hot boundary of the cooling zone has $T\sim 3\times 10^5\,$K,
the shear velocity of the well-developed cooling zone $v_{x,\text{max}}\sim
c_\text{s}(3\times 10^5\,\text{K})\sim 0.55c_\text{s,h}$.} 

\begin{figure}
    \centering
    \includegraphics[width=\columnwidth]{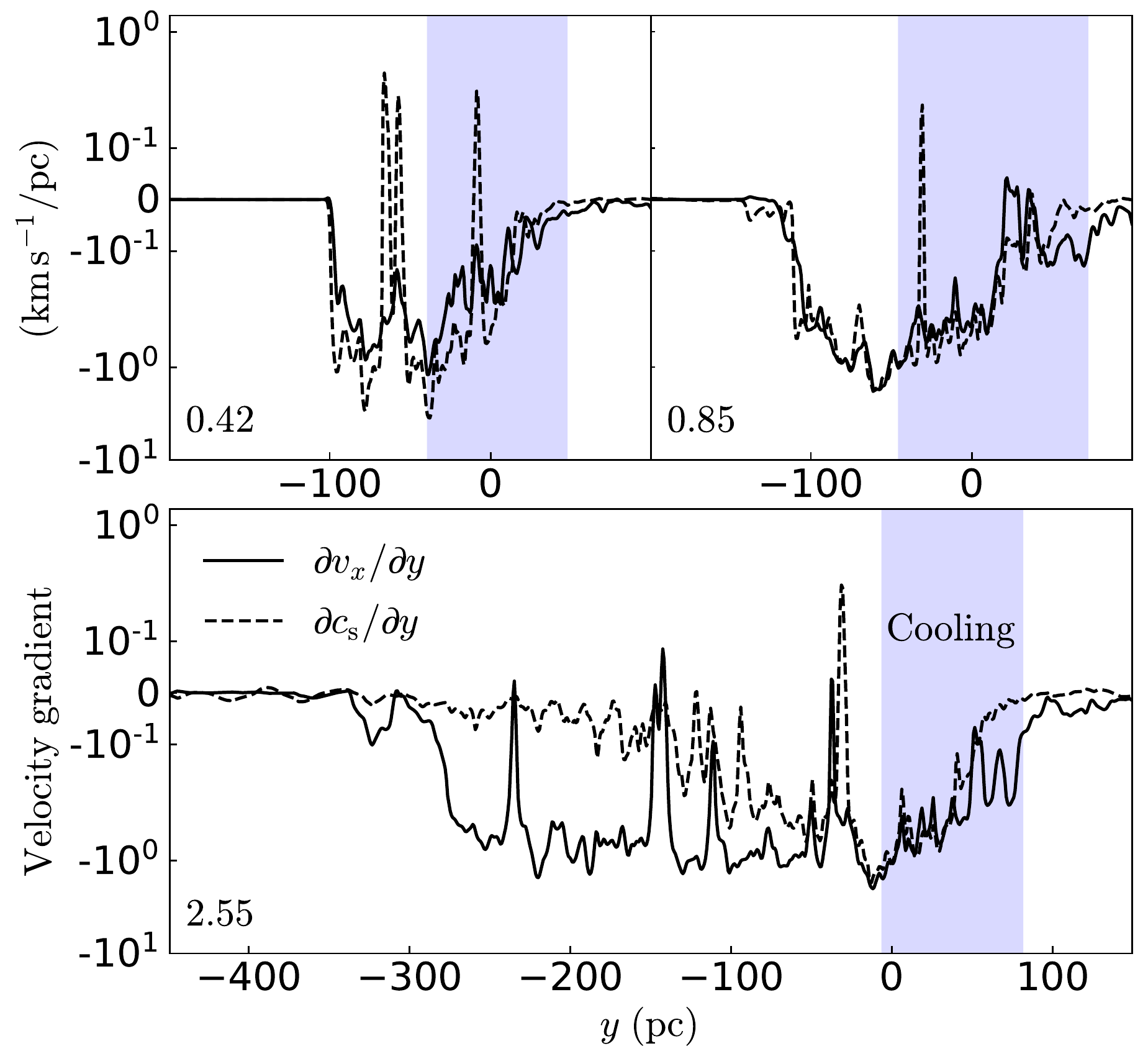}
    \caption{Profiles of the gradients of $x$-velocity (solid) and sound
    speed $c_\text{s}$ (dashed) in the saturation stage ($t=80\,$Myr), where regions in
    which cooling and mixing are significant are shadowed in blue. In the mixing
    regions, the gradients of $v_x$ and $c_\text{s}$ are almost identical over a
    factor of $10$ in their magnitudes, double-confirming their magnitude
    profiles closely trace each other in Fig.~\ref{fig:vx_compare}.}
    \label{fig:v_gradient}
\end{figure}

Although the problem we study inherently has Galilean invariance 
which makes velocity gradient rather than the magnitude of velocity 
matter, we directly plot velocity profiles firstly, for two reasons: 
(i) $v_{x,\mathrm{min}}=0$ in our setup and $c_\mathrm{s,min}$ is 
very small (we ultimately compare shapes of profiles that determine 
gradients, instead of magnitudes, and it makes sense to plot both 
profiles together if $v_x$ and $c_\mathrm{s}$ have almost the same 
minimum value) (ii) it is easier to compare runs with different Mach 
numbers in this concise plot than a plot of velocity gradients. In 
case the velocity gradients are of interest, equivalently, we also 
have them plotted in Fig.~\ref{fig:v_gradient} (three of the runs), 
which likewise shows that the gradient of $v_x$ (i.e., $\partial 
v_x / \partial y$) is approaching and then tracing $\partial c_\mathrm{s} / \partial y$ 
with increasing $\mathcal{M}$ in the mixing (cooling) zone while 
$\partial v_x / \partial y$ exceeds (in terms of magnitude) and 
diverge from $\partial c_\mathrm{s} / \partial y$ outside the cooling 
zone in high-$\mathcal{M}$ runs.

\subsection{Separation of turbulent and mixing zones}
\label{sec:division}

\begin{figure*}
    \centering
    \includegraphics[width=0.494\linewidth]{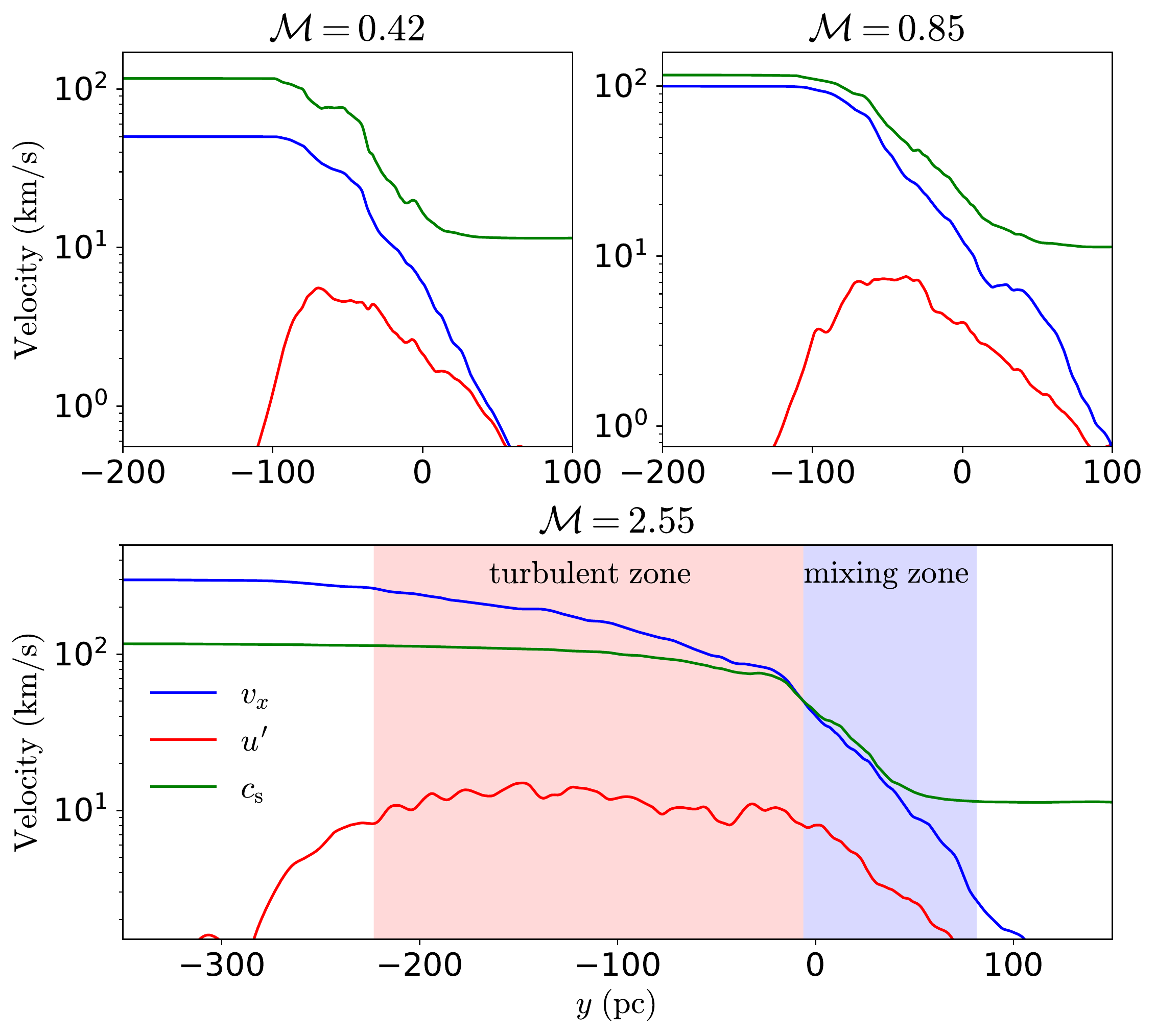}%
    \includegraphics[width=0.505\linewidth]{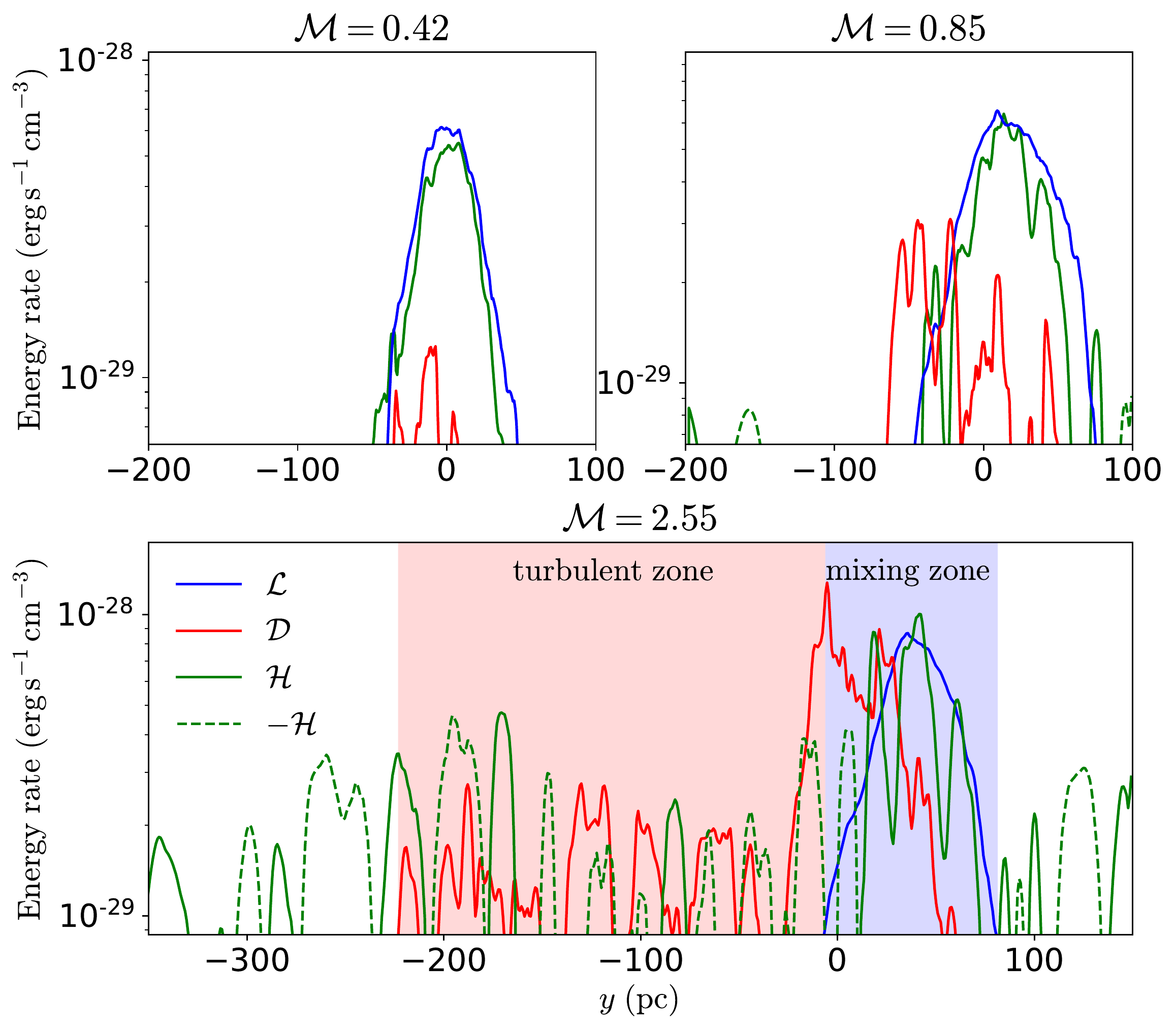}
    \caption{\textit{Left panel:} Profiles (along the $y$ axis) of the shear
        velocity $v_x$ (blue), turbulent velocity $u'$ (red) and local sound
        speed $c_\text{s}$ (green) in the saturation stage ($ t =
        80\,\mathrm{Myr}$) in simulations {\tt M0.42} (top-left), {\tt M0.85}
        (top-right) and {\tt M2.55} (bottom). \textit{Right panel:} Profiles of
        radiative cooling rate $\mathcal{L}$ (blue), turbulent dissipation rate
        $\mathcal{D}$ (red) and the divergence of enthalpy flux (i.e., enthalpy
        loss rate) $ \mathcal{H}$ (green) with $\mathcal{H}$ in solid and
        $-\mathcal{H}$ in dashed. At low Mach numbers ({\tt M0.42} and {\tt
        M0.85}), highly-turbulent regions and mixing regions are roughly
        co-spatial, where the divergence of enthalpy flux compensates radiative
        cooling in mixing regions. At high Mach numbers ({\tt M2.55}), the
        turbulent mixing layer is split into two separated zones: the mixing
        zone (blue shadowed) with strong cooling and predominant turbulent
        dissipation (the region in which $\mathcal{L}$ is not below
        $\mathcal{L}_\mathrm{max}/10$ in profile is defined as the mixing region
        here) , and the turbulent zone (red shadowed) with large turbulent
        velocities and weak turbulent dissipation (the boundaries of the
        turbulent zone here are determined by the ``hot'' end of the mixing zone
        and the point at which $\mathcal{D}$ reaches
        $\mathcal{D}_\mathrm{max}/10$ in profile).}
    \label{fig:turb_vx_lum}
\end{figure*}

In Fig.~\ref{fig:turb_vx_lum}, we show profiles of velocities (left) and energy
rates (right) in simulations {\tt M0.42}, {\tt M0.85} and {\tt M2.55}. In the
left panel are profiles of the bulk velocity $v_x$, turbulent velocity $u'$ and
the local sound speed $c_\text{s}$. In all cases, the shear velocity profiles
$v_x$ do not exceed the local sound speed $c_\text{s}$ in the cooling region
even in the supersonic run {\tt M2.55}, as already shown in
Fig.~\ref{fig:vx_compare} and double-confirmed here. We also explicitly verified
in all cases, that the turbulent velocity profiles are proportional to gradients
of $v_x$, which is expected by the mixing length theory mentioned in
\S\ref{sec:velocity_constraint}.\footnote{Though it is not so clear in the
figure with logarithmic scales (Fig.~\ref{fig:turb_vx_lum}), the proportionality
$u' \propto \partial v_x / \partial y$ is obvious in linear scales.} However,
the velocity profiles differ significantly between low and high Mach numbers: in
runs {\tt M0.42} and {\tt M0.85}, the highly turbulent regions are roughly
co-spatial with the mixing/cooling regions traced by intermediate sound speeds, as expected in traditional ``turbulent'' ``mixing'' layers. In
contrast, in the high-Mach number run {\tt M2.55}, the high-turbulent region
(shadowed in red) spanning over $200\,\mathrm{pc}$ develops \emph{outside} of
the mixing region (shadowed in blue), indicating that a turbulent mixing layer
(the entire turbulent region) at high Mach numbers is characterized by two
separated zones, the ``turbulent zone'' with high turbulent velocities, and the
``mixing zone'' with significant mixing and cooling.

The right panel of Fig.~\ref{fig:turb_vx_lum} shows profiles of the radiative
cooling rate $\mathcal{L}$ (blue), turbulent dissipation rate $\mathcal{D}$
(red) and enthalpy loss rate $\mathcal{H}$ (green, refer to
Table~\ref{tab:notation} for detailed definitions of these quantities). The
negative values $-\mathcal{H}$ are also plotted (dashed green), representing
enthalpy gain rate due to local gas expansion. In the run {\tt M0.42},
$\mathcal{H}$ and $\mathcal{L}$ overlap quite well, while $\mathcal{D}$ is
negligible. This indicates the enthalpy flux of hot gas is responsible for
balancing radiative cooling in the TML, which is consistent with previous
findings \citep{Ji2019}. In {\tt M0.85} where the shear flows get close to
transonic, $\mathcal{D}$ becomes significant, i.e., turbulent dissipation starts
to play a role in energizing the cooling zone. In these two subsonic runs,
$\mathcal{H}$ is still more important than $\mathcal{D}$ in balancing cooling
via radiation. The profiles with significant $\mathcal{H}$ but negligible
$-\mathcal{H}$ show that the diffuse hot gas tends to be compressed during the
process of mixing and radiative cooling, while gas expansion is not
considerable.

However, in the supersonic case {\tt M2.55} (bottom-right panel of
Fig.~\ref{fig:turb_vx_lum}), turbulent dissipation becomes the dominant energy
source compensating radiative cooling, while the net contribution from
$\mathcal{H}$ is very tiny (will be clearly shown in Fig.~\ref{fig:surf_Mach}).
Compared with the mixing zone, turbulent dissipation in the turbulent zone is
weaker, although the turbulent velocities in the turbulent zone are higher (as
shown in the left panel), which is due to lower densities of hot gas which carry
less turbulent energy. The enthalpy loss rate in the TML violently fluctuates
between positive and negative values, corresponding to the expansion and
compression of gas respectively. The sum of fluctuations in the enthalpy loss
rate in the entire turbulent region is much less than turbulent dissipation.

\begin{figure}
    \centering
    \includegraphics[width=\columnwidth]{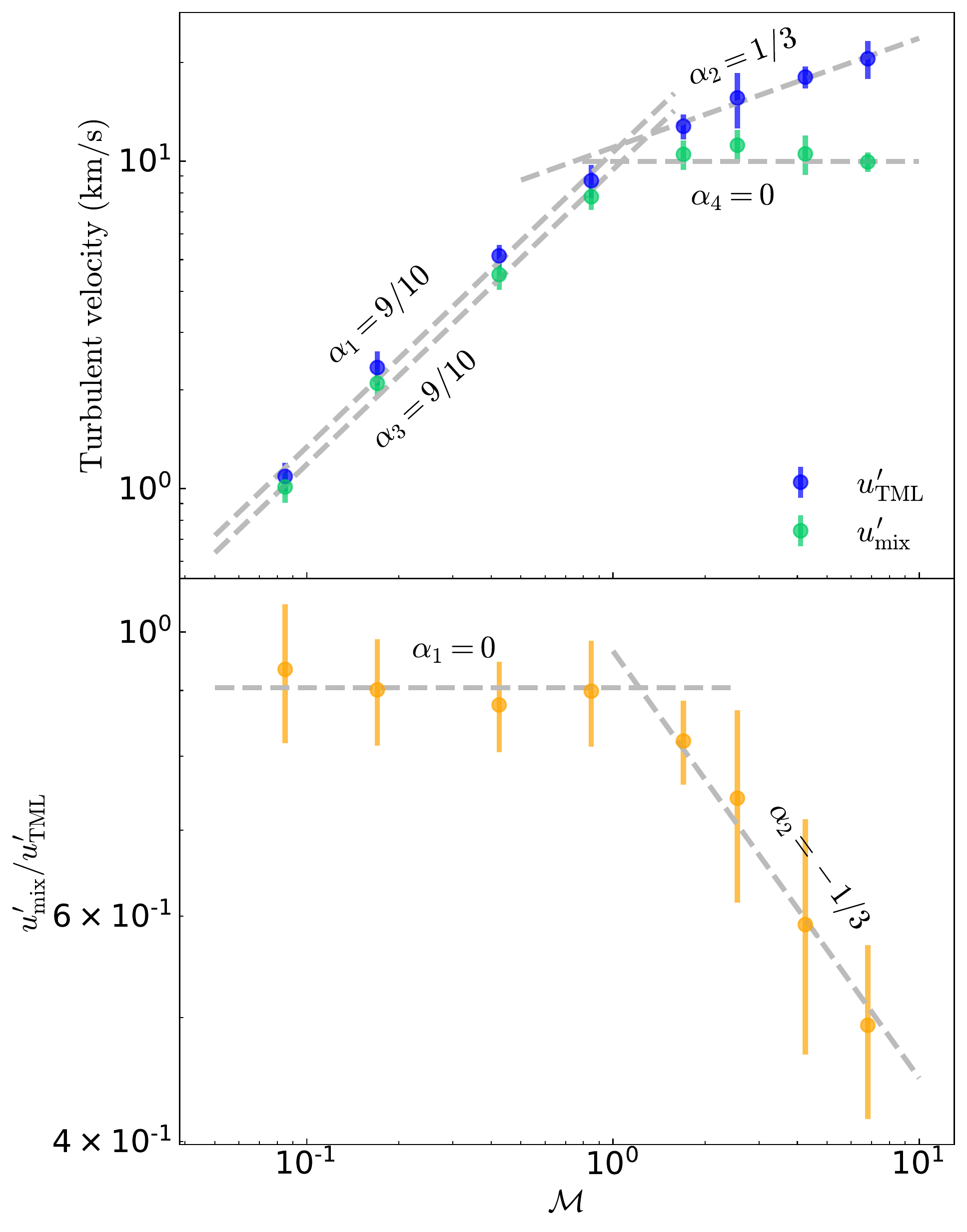}
    \caption{{\it Upper panel:} Turbulent velocities in the entire TML
    $u'_\mathrm{TML}$ (blue) and in the mixing zone only $u'_\mathrm{mix}$
    (green) plotted against Mach number. {\it Lower panel:} The ratio of
    $u_\text{mix}'$ to $u_\text{TML}'$ plotted against $\mathcal{M}$. At
    $\mathcal{M}\lesssim 1$, $u_\text{mix}'/u_\text{TML}'$ is independent of
    $\mathcal{M}$, suggesting a mixing zone, if exists, does not distinguish
    from the whole TML region. However, in cases of $\mathcal{M}\gtrsim 1$,
    $u_\text{mix}'$ and $u_\text{TML}'$ start to follow different power law
    relations, indicating that a mixing zone with different properties of
    turbulence stands out from the TML. Error bars in the figure and hereafter
    represent one standard deviation of uncertainty around the mean values,
    which are measured only when the simulations reach the saturation stage.}
    \label{fig:turb_Mach}
\end{figure}

The separation of the turbulent and the mixing zones at high Mach numbers is
more than a visual fact; their physical properties are intrinsically different.
In the top panel of Fig.~\ref{fig:turb_Mach}, we plot the Mach number dependence
of turbulent velocities, where the blue and green dots represent the peak
turbulent velocities in the entire turbulent region
$u'_\mathrm{TML}$\footnote{Since the turbulent zone does not remarkably
distinguish from the mixing layer within the entire turbulent region in the
subsonic case, we compute the turbulent velocity in the entire turbulent region
rather than that ``turbulent zone'' we define. As in \citet{Tan2021}, the peak
turbulent velocity is measured to characterize the TML.} and the local turbulent
velocities\footnote{We define the local (peak) turbulent velocity of the gas in
the mixing zone as the turbulent velocity of the gas enriched with \ion{O}{VI}
at $\sim 3\times 10^5\,$K (in practice, we pick up gas at $2$-$4\times
10^5\,$K).} in the mixing zone $u'_\mathrm{mix}$ respectively. At low Mach
numbers, both the TML and the mixing zone follow a similar power-law of
$\mathcal{M}^{9/10}$, i.e., their behaviors (dependencies on $\mathcal{M}$) are
parallel. However, this is not the case for high Mach numbers: the Mach number
dependence of $u'_\mathrm{TML}$ follows a power-law of $\mathcal{M}^{1/3}$ at
$\mathcal{M}\gtrsim 1$, i.e., the increase in shear velocities indeed enhances
turbulent motions in the turbulent region. In contrast, $u'_\mathrm{mix}$
becomes independent of $\mathcal{M}$ at high Mach numbers, indicating that
turbulent velocities in the mixing zone cease to increase further with larger
initial shear velocities. The bottom panel of Fig.~\ref{fig:turb_Mach} plots the
ratio of the turbulent velocity in the mixing zone to that in the entire TML
$u'_\mathrm{mix}/u'_\mathrm{TML}$, versus $\mathcal{M}$ in each simulation. The
ratio is independent of $\mathcal{M}$ at low Mach numbers, while follows
$\mathcal{M}^{1/2}$ at high Mach numbers, further implying that the behaviors of
the entire turbulent region and the mixing zone diverge for $\mathcal{M}\gtrsim
1$. Consequently, in the supersonic case, the turbulent region grows larger with
greater $u'_\text{TML}$ at higher $\mathcal{M}$, but the mixing zone does not
with an approximately constant $u'_\text{mix}$, and the mixing zone and
turbulent zone become distinguishable.

\subsection{Balance of Energies}
\label{sec:energies}

We next focus on the volume-integral quantities of turbulent mixing layers from
the perspective of energetics. In the case that $\mathcal{M}\lesssim 1$
\citep{Ji2019}, as Eq.~\eqref{eq:balance} indicated, a quasi-equilibrium state,
i.e., the statistical properties of TMLs do not evolve further with time,
is reached when the sum of enthalpy flux and kinetic energy flux balances the
energy loss via radiative cooling. During the process of energy transporting and
transforming, hot gas carrying enthalpy and kinetic energy is entrained by
turbulence into the TML. In the TML, some portion of the kinetic energy is
converted to heat through turbulent dissipation, and the enthalpy is consumed through radiative cooling (accompanied by gas compression), and finally cools radiatively.
The process above in the turbulent region in a quasi-equilibrium state can be
described as
\begin{equation}
    F_\text{h+k}\sim D+H \sim Q,
    \label{eq:energy_process}
\end{equation}
where $F_\mathrm{h+k}$ is the flux of enthalpy and kinetic energy (into the
turbulent region from outside), $D$ the turbulent dissipation and $H$ the enthalpy
loss (see Table~\ref{tab:notation} for detailed definitions which are derived
from time-independent hydrodynamic conservation equations,
\citealt{Favre1969,Kuncic2004}).\footnote{$D+H=Q$ is literally the integral form
of the energy equation from the time-independent hydrodynamic equations,
assuming the sum of gas internal energy in the turbulent region does not change,
$-t_{ij}^\text{R}\tilde{S}_{ij}-\bar{P}\tilde{v}_{i,i}-\left<Pv_{i,i}' \right> =
n^2 \Lambda$ (in the frame where the cold-hot interface is static -- see
Table~\ref{tab:notation} for detailed notations).} However, we note that
Eq.~(\ref{eq:energy_process}) is only a necessary but not sufficient condition for an
equilibrium state, which further requires that
\begin{equation}
    \left\{
    \begin{array}{l}
        F_\text{h} \sim H \\
        F_\text{k} \sim D
    \end{array},
    \right.
    \label{eq:fine_balance}
\end{equation}
i.e., the enthalpy flux $F_\text{h}$ replenishes the enthalpy of the TML, while
the kinetic energy flux $F_\text{k}$ makes up for the kinetic energy consumed by
dissipation. This requirement is self-evident by definition: a quasi-equilibrium
TML which does not evolve further with time should maintain a constant turbulent
dissipation $D$. Therefore, the kinetic energy flux into TMLs, as the only
energy source of turbulent dissipation, should be a time-independent constant as
well with $F_\text{k}\sim D$. Following Eq.~\eqref{eq:energy_process},
$F_\text{h} \sim H$ is thus needed for a quasi-equilibrium state as well.

\begin{figure*}
    \centering
    \includegraphics[width=\textwidth]{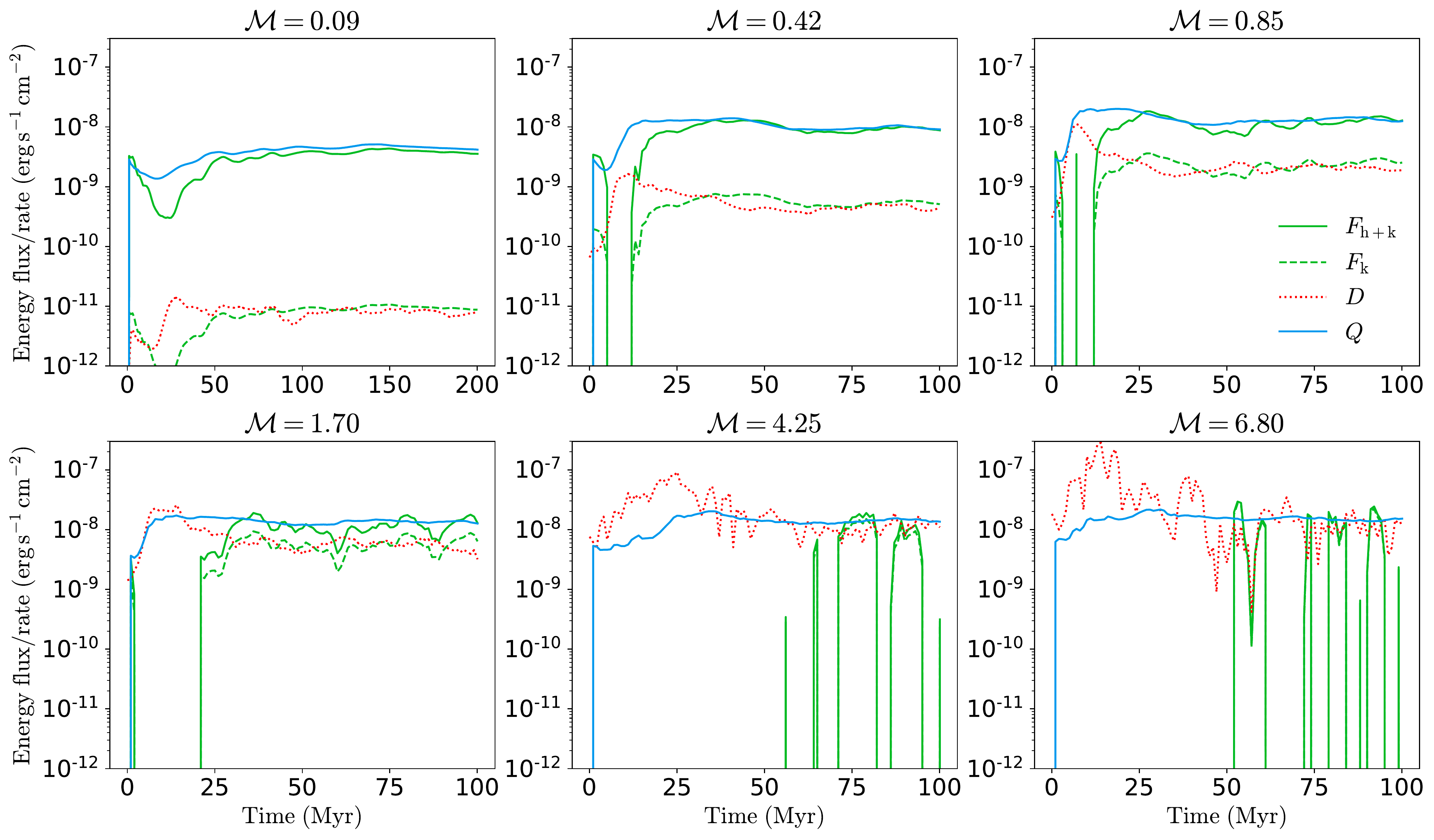}
    \caption{Time evolution of selected mean energy fluxes and rates from
    simulations with different Mach numbers: the sum of enthalpy flux
    and bulk kinetic energy flux $F_\text{h+k}$ (green-solid), the flux of bulk
    kinetic energy $F_\text{k}$ (green-dashed), the averaged dissipation rate
    over the whole turbulent region $D$ (red-dotted) and the surface brightness
    of the TML $Q$ (blue-solid). In subsonic cases, TMLs ultimately reach a
    quasi-equilibrium state with $F_\text{h+k}\sim Q$ and $F_\text{k}\sim D$,
    where turbulent dissipation $D$ is negligible compared to $Q$. In supersonic
    cases, however, turbulent dissipation $D$ becomes important and dominant;
    $F_\text{h+k}\lesssim Q$ for transonic Mach numbers ({\tt M1.70}) and
    $F_\text{h+k}\ll Q$ in highly supersonic runs ({\tt M4.25}, {\tt M6.80}),
    indicating TMLs are not in quasi-equilibrium. Note: The negative fluxes (not shown on
    logarithmic scales) during the early stage are caused by the initial
    expansion of the turbulent region (significant turbulent pressure), which is
    transient and does not imply a steady flux outwards from TMLs.}
    \label{fig:energy_flux}
\end{figure*}

To examine how well the energy balance is maintained, the time evolution of some
selected energy fluxes/rates from our simulations are plotted in
Fig.~\ref{fig:energy_flux}. Consistent with theoretical predictions above, in
subsonic cases ($\mathcal{M}\lesssim 1$, top panels of
Fig.~\ref{fig:energy_flux}), the sum of enthalpy and bulk kinetic energy flux
into the turbulent mixing layer $F_\text{h+k}$ well balances radiative cooling
$Q$, and the TML reaches quasi-equilibrium described by Eqs.~\eqref{eq:balance}
and~\eqref{eq:energy_process}. The contribution from kinetic energy $F_\text{k}$
to energizing radiative TMLs is modest compared to enthalpy $F_\text{h}$, i.e.,
the internal energy transported by gas advection and compression is much more
significant than turbulent dissipation in balancing radiative cooling.

However, the case is different at high Mach numbers. As shown in the bottom
panels of Fig.~\ref{fig:energy_flux}, {\tt M1.70} only marginally reaches
quasi-equilibrium ($F_\text{h+k}\lesssim Q$, $F_\text{k}\sim D$), whereas {\tt
M4.25} and {\tt M6.80} are far from quasi-equilibrium with $F_\text{h+k}\ll
Q\sim D$, even when the surface brightness $Q$ is still steady with time. In
other words, at high Mach numbers, the mixing zone of the TML is indeed in a
steady state characterized by an approximately constant $Q$, but the relative
advection of hot gas to the TML is almost quenched. This non-equilibrium state
actually corresponds to the evaporation\footnote{The non-equilibrium state is
intuitively expected in the case of evaporation, since a flow of ``hot gas
condensing into cold gas'' in equilibrium cannot form when the cold phase is
being continuously absorbed and evaporated by the TML which causes a continuous
growth of the TML.} of cold gas, which will be discussed in
\S\ref{sec:cold_gas_mass}.

\begin{figure}
    \includegraphics[width=\columnwidth]{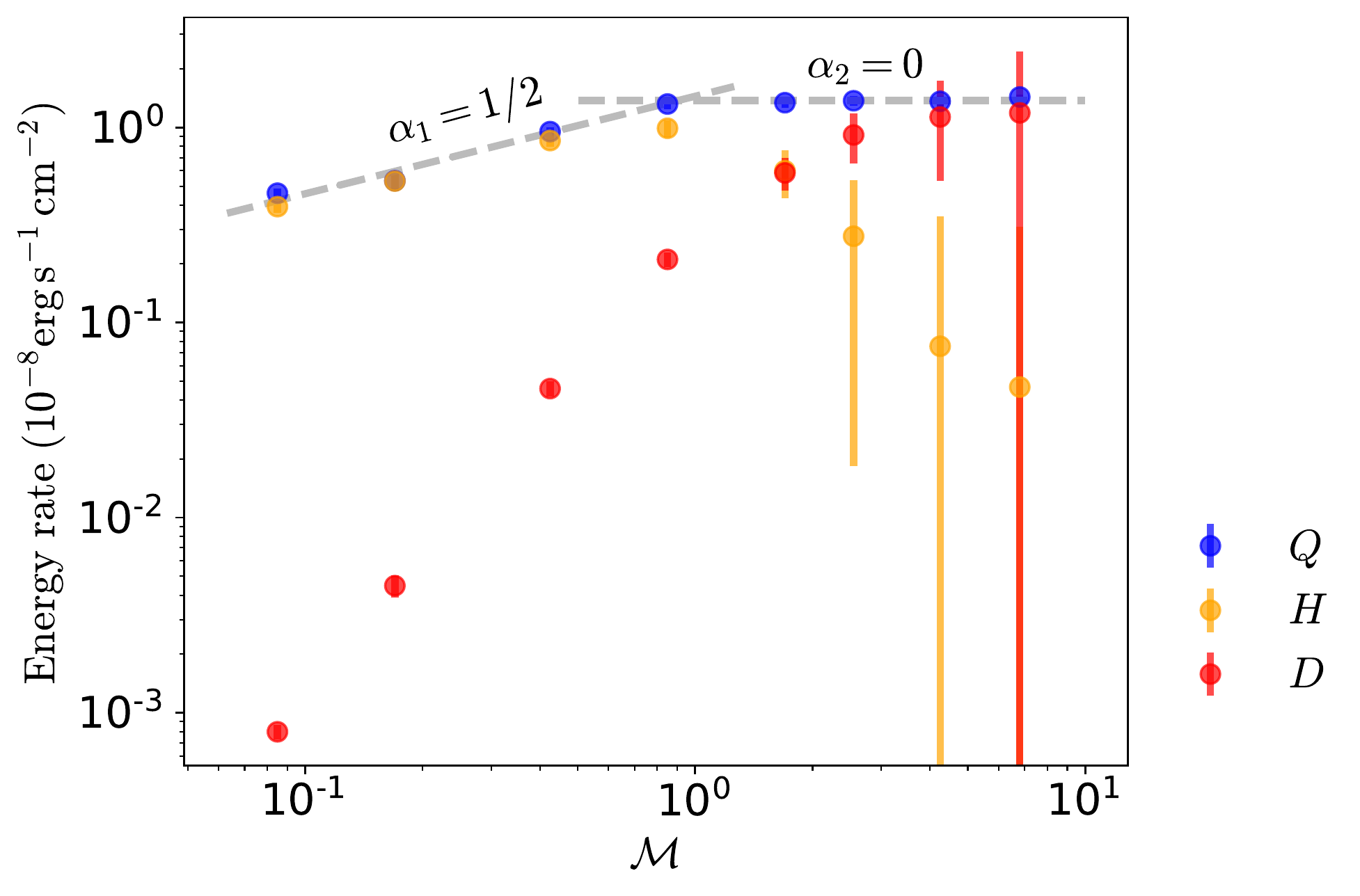}
    \caption{The surface brightness $Q$ (blue), enthalpy consumption $H$
    (orange), and turbulent dissipation $D$ (red) plotted against Mach number.
    At low Mach numbers, TMLs follow $Q\propto \mathcal{M}^{1/2}$ with the
    enthalpy as the primary energy source. However, the surface brightness $Q$
    becomes independent of $\mathcal{M}$ at high Mach numbers with increasingly
    important turbulent dissipation $D$ rather than $H$ that energizes TMLs.}
    \label{fig:surf_Mach}
\end{figure}

Fig.~\ref{fig:surf_Mach} further shows the surface brightness $Q$, turbulent
dissipation $D$ and enthalpy loss $H$ taken from simulations with different Mach
numbers, plotted against $\mathcal{M}$ of each simulation. The surface
brightness $Q$ scales with $\mathcal{M}^{1/2}$ for $\mathcal{M}\lesssim 1$
(consistent with previous findings, e.g., \citealt{Ji2019}), and saturates for
$\mathcal{M}\gtrsim 1$. At low Mach numbers, the enthalpy loss $H \sim Q$ and
the turbulent dissipation $D$ is negligible, indicating that low-Mach number
TMLs are energized by enthalpy flux. In contrast, at high Mach numbers,
turbulent dissipation becomes increasingly dominant with $D\sim Q \gg H$.
Although previous studies  \citep{Ji2019} have found turbulent dissipation can
become non-negligible with stronger cooling, it is not as predominantly
important as the supersonic conditions studied here.

In short, regarding energy balancing, TMLs are distinctively different at low
and high Mach numbers. At low Mach numbers, the TML surface brightness scales with
$\mathcal{M}^{1/2}$. Low-Mach number TMLs are energized by enthalpy flux
$F_\text{h}$, and can reach quasi-equilibrium. In contrast, at high Mach
numbers, the surface brightness saturates and becomes independent of
$\mathcal{M}$, and the dominant energy source is turbulent dissipation
(sourced from $F_\text{k}$) rather than enthalpy. And a quasi-equilibrium
state cannot be reached in high-Mach number TMLs.

\subsection{Scaling relations}
\label{sec:scalings}

In this section, we summarize the scaling relations of a few important TML
quantities relative to Mach number $\mathcal{M}$:

\begin{itemize}

    \item{\it Turbulent velocity} As already shown in Fig.~\ref{fig:turb_Mach},
    the turbulent velocity in the entire TML $u_\text{TML}'$
    follows the power law (not exact but a fairly good fit):
    \begin{equation}
        u_\text{TML}' = \left\{
        \begin{array}{l l}
        u_\text{TML,1}' \mathcal{M}^{9/10} & \mathcal{M}\lesssim 1 \\
        \\
        u_\text{TML,2}' \mathcal{M}^{1/3}  & \mathcal{M}\gtrsim 1
        \end{array},
        \right.
        \label{eq:v_turb_M}
    \end{equation}
    where $u_\text{TML,1}'$ and $u_\text{TML,2}'$ are normalization factors with
    values of $u_\text{TML,1}'\sim 10.64\,\mathrm{km/s} $ and $u_\text{TML,2}'\sim 11.01\,\mathrm{km/s}$ measured from our simulations. 
    
    Regarding the turbulent velocity in the mixing zone $u_\text{mix}'$
    where most mixing and cooling occur, Fig.~\ref{fig:turb_Mach} shows:
    \begin{equation}
        u_\text{mix}' = \left\{
        \begin{array}{l l}
      u_\text{mix,1}' \mathcal{M}^{9/10} & \mathcal{M}\lesssim 1 \\
      \\
      u_\text{mix,2}' \mathcal{M}^0   & \mathcal{M}\gtrsim 1
        \end{array},
        \right.
        \label{eq:v_turb_mix_M}
    \end{equation}
    where $u_\text{mix,1}'\sim 9.30\,\mathrm{km/s}$ and $u_\text{mix,2}'\sim 10.48\,\mathrm{km/s}$. 
    
    As already extensively discussed in \S\ref{sec:division}, both
    $u_\text{TML}'$ and $u_\text{mix}'$ obey the same power law $\propto
    \mathcal{M}^{9/10}$ when $\mathcal{M}\lesssim 1$, implying the mixing zone
    we defined here is indistinguishable from the entire TML region. In the
    supersonic case ($\mathcal{M} > 1$), however, unlike
    $u_\text{TML}'$, $u_\text{mix}'$ becomes approximately independent of $\mathcal{M}$. Again, this suggests that at high Mach numbers, the behaviors of the TML and
    the mixing zone start to diverge.

    \item{\it Surface brightness} As a direct indicator of the hot gas
    entrainment rate, the surface brightness $Q$ is among the key quantities of
    radiative TMLs. As shown in Fig.~\ref{fig:surf_Mach}, the scaling relation
    for $Q$ is:
    \begin{equation}
        Q \propto \left\{
        \begin{array}{l l}
        \mathcal{M}^{1/2} & \mathcal{M}\lesssim 1 \\
        \\
        \mathcal{M}^{0} & \mathcal{M}\gtrsim 1
        \end{array}.
        \right.
        \label{eq:Q_M}
    \end{equation}
    Combining this relation with Eq.~\eqref{eq:v_turb_M}, we have $Q\propto
    u_\text{TML}'^{5/9}$ for the subsonic case, which is virtually consistent with
    the result of $Q\propto u'^{1/2}$ obtained in \citet{Tan2021} with weak
    cooling. But for the supersonic case, $Q$ no longer increase with $\mathcal{M}$, though $u_\text{TML}'$ can be enhanced by a larger $\mathcal{M}$.

    \begin{figure}
        \centering
        \includegraphics[width=\columnwidth]{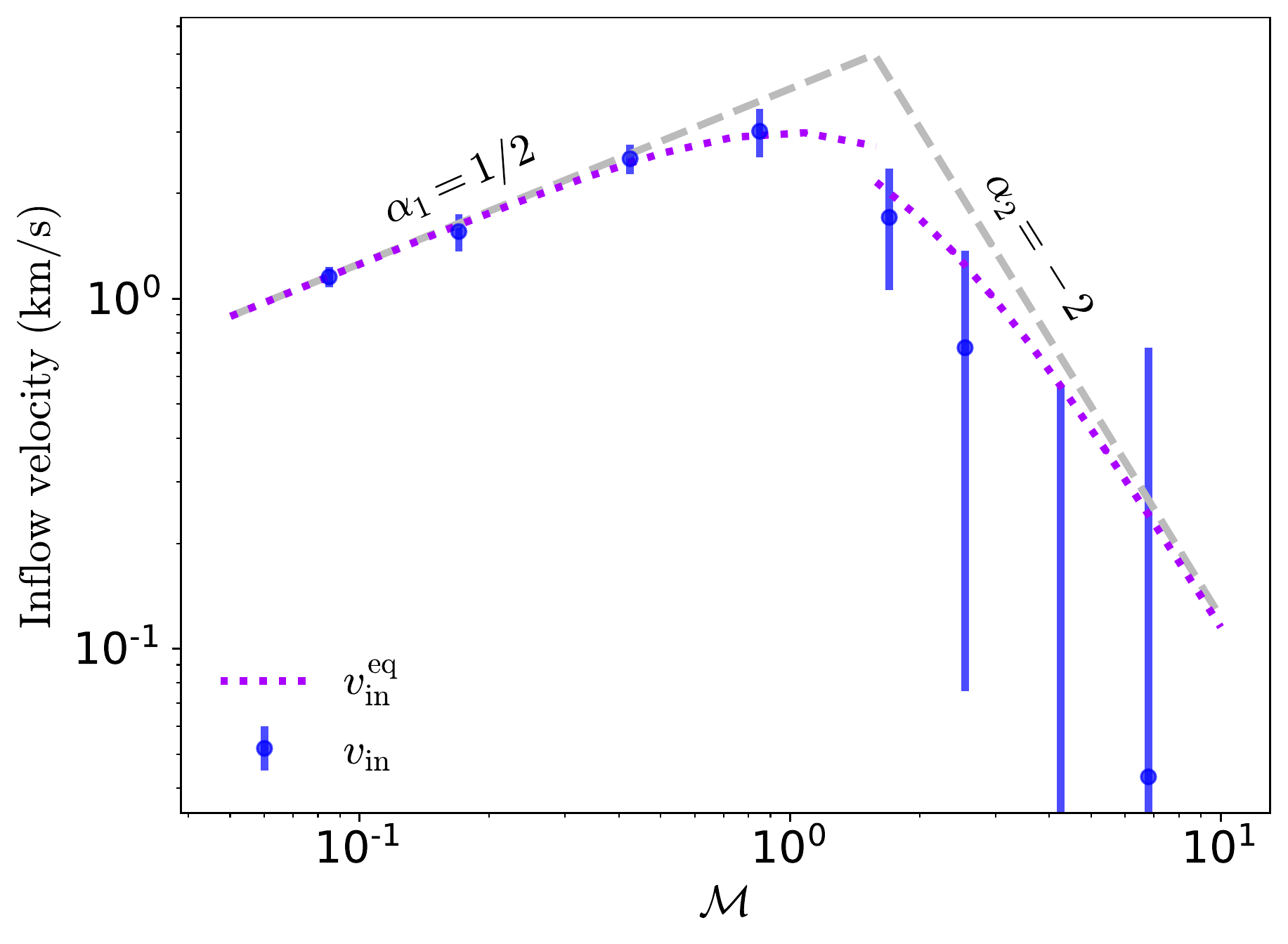}
        \caption{Inflow velocity into the TMLs $v_\text{in}$ (blue dots) plotted
        against Mach numbers $\mathcal{M}$. The dotted purple lines are the
        equilibrium inflow velocity $v_\text{in}^\text{eq}$ computed by
        Eq.~\eqref{eq:v_int_Mach}, with the assumption of a quasi-equilibrium
        state described by Eq.~\eqref{eq:balance}, and the dashed-grey lines are
        the asymptotes of $v_\text{in}^\text{eq}$ for $\mathcal{M}\rightarrow 0
        $ and $\mathcal{M}\rightarrow \infty$ respectively. The inflow velocity
        $v_\text{in}$ increases with $\mathcal{M}$ in the subsonic case, but
        decreases with $\mathcal{M}$ in the supersonic case. This indicates the
        entrainment of hot gas relative to the TML is significantly suppressed
        at high Mach numbers, even when the TML surface brightness still remains
        constant. The Mach number dependence of $v_\text{in}$ is well predicted
        by $v_\text{in}^\text{eq}$ at low Mach numbers, while falls below
        $v_\text{in}^\text{eq}$ at high Mach numbers, suggesting high-Mach
        number TMLs cannot reach a quasi-equilibrium state.}
        \label{fig:v_in_Mach}
        \end{figure}

    \item{\it Inflow velocity} We finally investigate the scaling relation of
    the inflow velocity $v_\text{in}$, which is calculated in a reference frame
    where the mixing layer is stationary and is shown in Fig.~\ref{fig:v_in_Mach} against $\mathcal{M}$. The quasi-equilibrium inflow velocity
    $v_\text{in}^\text{eq}$ is also plotted which is obtained by combining
    Eq.~\eqref{eq:balance} (which assumes a quasi-equilibrium state) and the
    dependency of $Q$ on $\mathcal{M}$, and has the following form:
    \begin{equation}
        v_\text{in}^\text{eq} \propto \left\{
        \begin{array}{l l}
            \displaystyle \frac{\mathcal{M}^{1/2}}{3+\mathcal{M}^2} & \mathcal{M}\lesssim 1\\
            \\
            \displaystyle \frac{1}{3+\mathcal{M}^2} & \mathcal{M}\gtrsim 1
        \end{array}.
        \right.
        \label{eq:v_int_Mach}
    \end{equation}
    We find that at low Mach numbers, the curve of the quasi-equilibrium inflow
    velocity $v_\text{in}^\text{eq}$ well predicts the inflow velocity
    $v_\text{in}$, suggesting that a quasi-equilibrium is reached with
    $\mathcal{M}\lesssim 1$. In particular, at $\mathcal{M}\ll 1$, we have
    $v_\text{in} \propto \mathcal{M}^{1/2}$, which has the same Mach number
    dependence with the surface brightness $Q$. This further demonstrates that
    low-Mach number TMLs are primarily energized by the enthalpy of hot gas
    entrained at an inflow velocity $v_\text{in}$. However, at high Mach
    numbers, the inflow velocity into the TML $v_\mathrm{in}$ decreases with
    increasing $\mathcal{M}$, indicating hot gas entrainment into the TMLs is
    significant \emph{suppressed} under supersonic conditions, even when the TML
    surface brightness is still comparable with that at transonic Mach numbers (see,
    e.g., Fig.~\ref{fig:surf_Mach}). This is consistent with the energy budget
    where the enthalpy flux from hot gas entrainment becomes
    subdominant at greater $\mathcal{M}$ in high-Mach number TMLs
    (\S\ref{sec:energies}). Moreover, $v_\text{in}$ falls below
    $v_\text{in}^\text{eq}$, double-confirming that a quasi-equilibrium state
    \emph{cannot} be reached for high-Mach number TMLs (as previously discussed
    in \S\ref{sec:energies}).

\end{itemize}

\subsection{Mass evolution of cold gas}
\label{sec:cold_gas_mass}

As previously discussed, the anti-correlation between $v_\mathrm{in}$ and
$\mathcal{M}$ at high Mach numbers indicates that the turbulent zone inhibits
the hot gas flowing into the TML. Therefore, instead of absorbing the hot gas,
in this case, it is the cold gas that is mixed into the TML and evaporates due
to strong turbulent dissipation. This is in contrast to the low Mach number case
where the hot gas is efficiently entrained and then condenses into the cold gas,
and thus the mass of cold phase grows with time. Moreover, at large
$\mathcal{M}$, since there always exists a net mass influx into TMLs contributed
by primarily evaporated cold gas plus (suppressed) entrained hot gas, high-Mach
number TMLs cannot reach a quasi-equilibrium state with both the mass and size
of TMLs growing with time. But in low-Mach number TMLs, the hot gas entrained
into TMLs ultimately condenses into the cold phase through TMLs, and thus TMLs
maintain a relatively fixed size and mass, and reach quasi-equilibrium.

\begin{figure}
    \centering
    \includegraphics[width=\columnwidth]{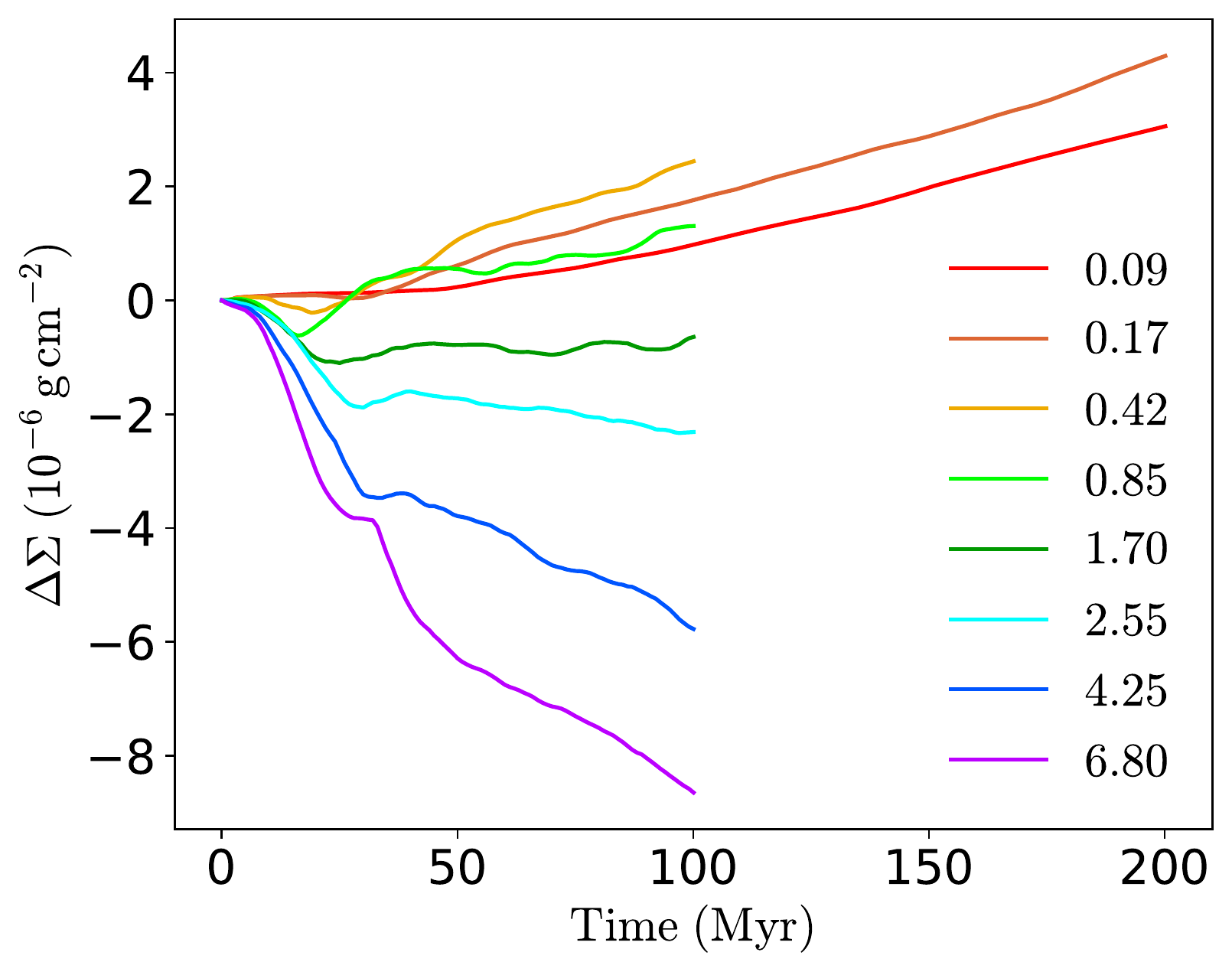}
    \caption{Mass evolution of cold gas in runs with different Mach numbers
    represented by colors, where $\Delta \Sigma$  is the change in column mass
    density of the cold gas. The cold gas mass grows with time in low-Mach
    number runs ($\mathcal{M}\lesssim 1$), while decreases with time in
    high-Mach number runs with evaporation rates increasing with greater
    $\mathcal{M}$.}
    \label{fig:cold_mass}
\end{figure}

Quantitatively, Fig.~\ref{fig:cold_mass} shows the time evolution of the cold
gas column mass density in our simulations, where the colors represent different
Mach numbers adopted in each simulation. When calculating the change of the cold
gas mass, the increase/decrease caused by the inflowing/outflowing cold gas
crossing the top box boundary has been subtracted, so our calculation accurately
reflects the cold gas mass evolution caused by TML physics only. The cold gas
mass in low-Mach number runs ({\tt M0.09}, {\tt M0.42} and {\tt M0.85}) appears
to grow with time due to the condensation of hot gas into the cool phase.
However, at high Mach numbers ({\tt M1.70}, {\tt M2.55}, {\tt M4.25} and {\tt
M6.80}), the cold gas mass decreases with evaporation rates increasing with
greater Mach numbers. The implication for the cloud-crushing problem is that the
cloud-wind interfaces can gain mass from hot gas entrainment and condensation,
and the cloud tends to survive in a hot wind with lower $\mathcal{M}$. But in a
higher-Mach number hot wind, when the hot gas entrainment into TMLs is
suppressed, the cloud could lose mass from the cloud-wind interfaces due to cold
gas evaporation. The tendency of cold gas mass evolution relative to the Mach
number is qualitatively consistent with cloud-crushing simulations (e.g.,
\citealt{Gronke2018}).

We finally end this subsection with a closing remark. Our finding that higher
Mach numbers lead to faster mass loss of cold gas in TMLs might sound intuitive
and unsurprising. However, this is actually not straightforward on second
thought, since based on theories of linear analysis, the KH instability is
suppressed under high-$\mathcal{M}$ conditions, which in principle facilitates
the survival and even the growth of cold gas. However, the linear analysis does
not deal with the long-term evolution at later times, when the velocity shear in
initial conditions smooths out due to finite diffusivity and the evolution with
radiative cooling becomes non-linear. Focusing on the long-term evolution, our
study presents a physical picture where the cold gas actually suffers from mass
loss via high-Mach number radiative TMLs. In this picture, the inflow velocity
and hot gas entrainment into TMLs are indeed suppressed, and this effect
actually cuts down further supplement of energy and mass from the hot gas to the
TMLs and the cold phase respectively. In the meanwhile, strong turbulent
dissipation (the energy is originally from the kinetic energy carried by the
hot gas\footnote{An equivalent statement given Galilean invariance could be: the
energy is originally from the bulk motion of the hot phase \textit{relative} to
the cold phase.} entrained into the TML at early times) which powers TMLs
evaporates the cold phase. This picture cannot be naively described as larger
velocities leading to a stronger KH instability and thus more efficient cold gas
mixing in high-Mach number TMLs.

\subsection{Ion column densities}
\label{sec:ion_column_dens}

\begin{figure*}
    \centering
    \includegraphics[width=\textwidth]{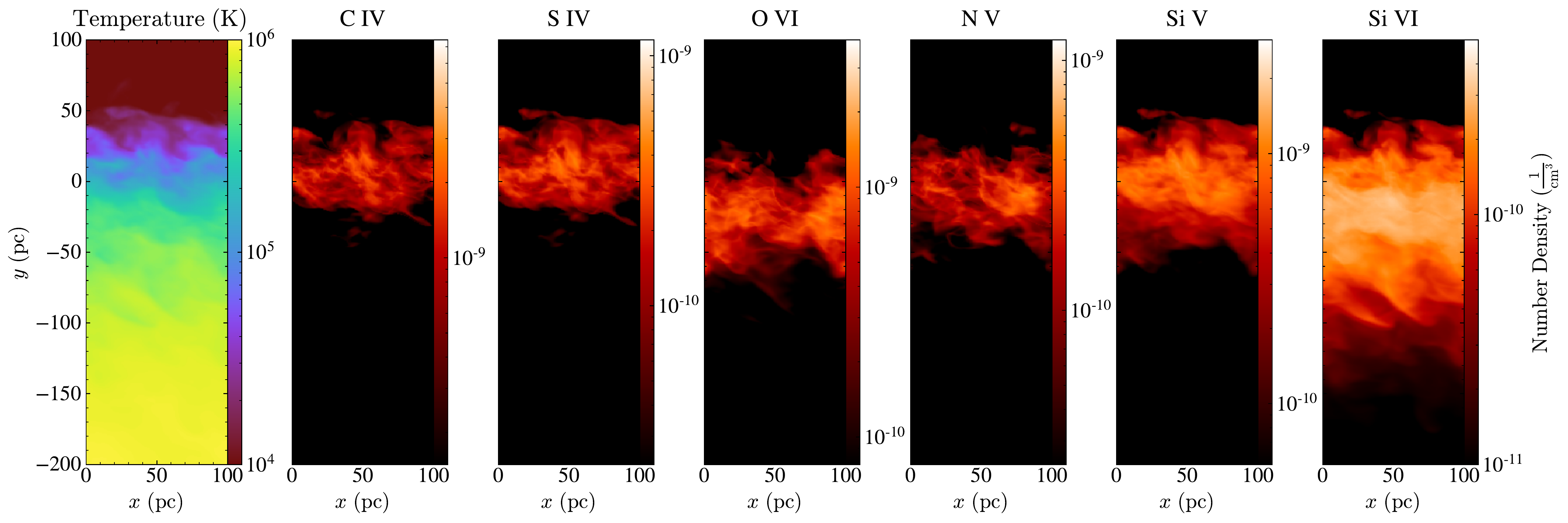}
    \caption{Volume-weighted projections of temperature and number densities of
        selected typical ions, \ion{C}{IV}, \ion{S}{IV}, \ion{O}{VI},
        \ion{N}{V}, \ion{Si}{V} and \ion{Si}{VI}, in {\tt M1.70} at $t=80\,$Myr.
        These ions closely trace the morphology of TMLs with slightly different
        preferred temperatures and densities.}
    \label{fig:ion_map}
\end{figure*}

\begin{figure}
    \centering
    \includegraphics[width=\columnwidth]{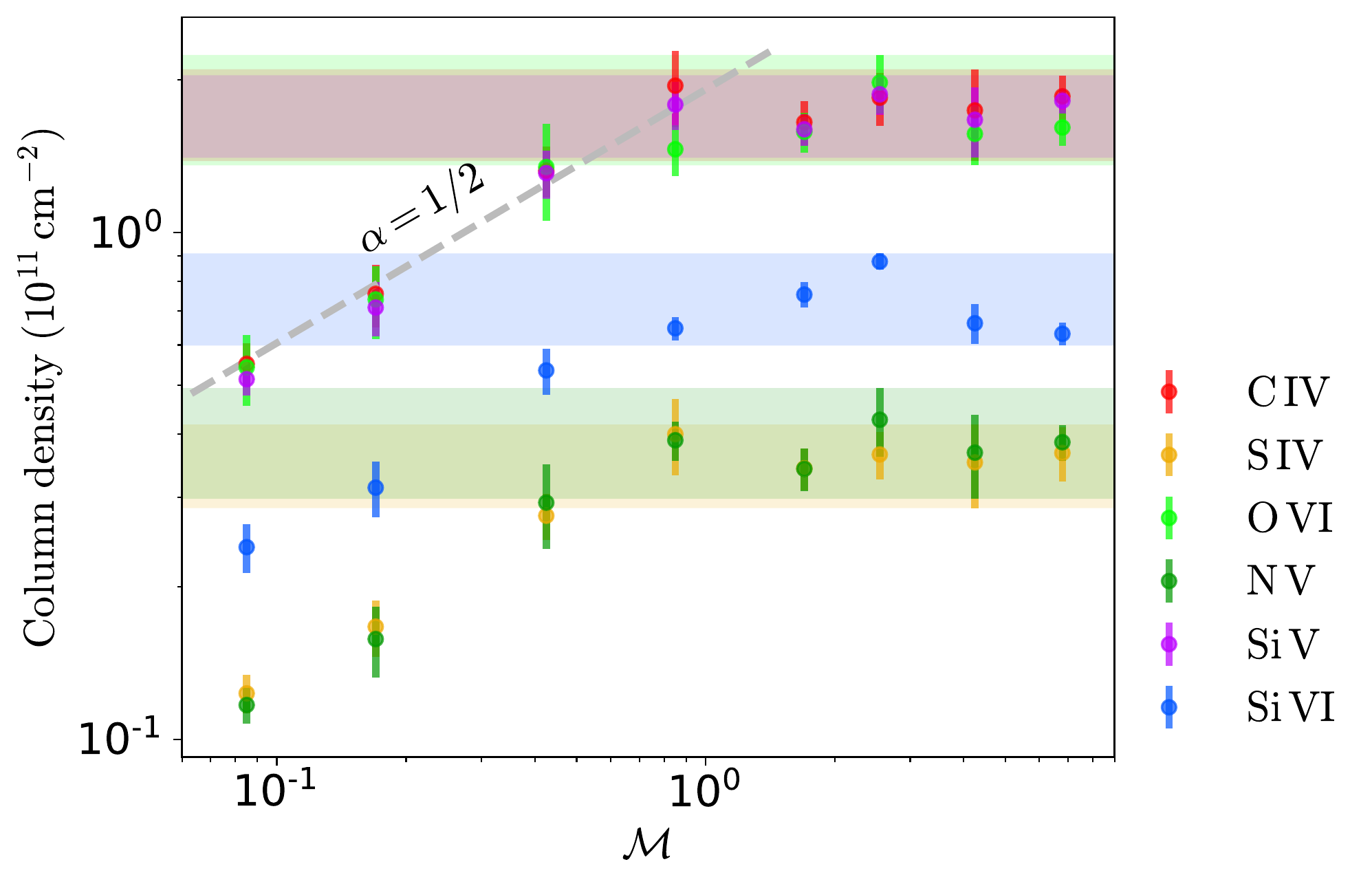}
    \caption{Column densities of \ion{C}{IV}, \ion{S}{IV}, \ion{O}{VI},
        \ion{N}{V}, \ion{Si}{V} and \ion{Si}{VI} plotted against Mach number,
        with $Z = 0.1 Z_\odot$ assumed. The column densities in subsonic runs
        increase with Mach number for thicker mixing layers, while they appear
        to be independent of Mach number in the supersonic case with saturated
        values (shadowed in colors): $N_\text{\ion{C}{IV}}\sim 1.77\times
        10^{11}\,\text{cm}^{-2}$, $N_\text{\ion{Si}{V}}\sim 3.56\times
        10^{10}\,\text{cm}^{-2}$, $N_\text{\ion{O}{VI}}\sim 1.68\times
        10^{11}\,\text{cm}^{-2}$, $N_\text{\ion{N}{V}}\sim 3.81\times
        10^{10}\,\text{cm}^{-2}$, $N_\text{\ion{Si}{VI}}\sim 7.32\times
        10^{10}\,\text{cm}^{-2}$, $N_\text{\ion{Si}{V}}\sim 1.74\times
        10^{11}\,\text{cm}^{-2}$. \textit{Note:} All ion column densities have
        been corrected accordingly so that they have identical initial values at
        different Mach numbers, by cutting off the extra column densities
        brought by longer boxes (high Mach-number runs), i.e., only the effects
        of radiative TMLs are under study (despite that the contribution from
        initial conditions are relatively modest).}
    \label{fig:column_Mach}
\end{figure}

We now examine the column densities of typical ions rising from the TMLs in the
simulations, where the number density of ions are computed with the assumption
of photoionization and collisional ionization equilibrium.
Fig.~\ref{fig:ion_map} shows the volume-weighted projections the number
densities of \ion{C}{IV}, \ion{S}{IV}, \ion{O}{VI}, \ion{N}{V}, \ion{Si}{V} and
\ion{Si}{VI} in {\tt M1.70} ($t = 80\,$Myr), as well as the corresponding
temperature projection. It is shown that these ions closely trace
intermediate-temperature gases, which significantly contribute to the mixing
layer surface brightness $Q$. Therefore, it is natural to expect that the
saturated ion column densities would follow a power law similar to $Q$ which
scales as $\mathcal{M}^{0.5}$. Indeed, the steady-state ion column densities do
not increase further with Mach number for $\mathcal{M}\gtrsim 1$, which is
clearly shown in Fig.~\ref{fig:column_Mach}: the ion column densities $N$
follow power laws of $N\propto \mathcal{M}^{0.5}$ for $\mathcal{M}\lesssim 1$
and $N\propto \mathcal{M}^0$ for $\mathcal{M}\gtrsim 1$. This is similar to the
Mach number dependence of surface brightness $Q$ (Fig.~\ref{fig:surf_Mach}) as
well.

At low Mach numbers, the magnitudes of ion column densities obtained in our
simulations are consistent with previous studies (e.g., \citealt{Ji2019}).
\citet{Ji2019} explored a range of cooling strengths, gas pressures and
metallicities, as well as time-dependent cooling and non-equilibrium ionization,
and concluded that under plausible CGM conditions, sightlines penetrating
hundreds or thousands of TMLs are required to explain the observed high column
densities (e.g., \ion{O}{VI}). Our findings further indicate that due to the
saturation at high Mach numbers, the maximum ion column densities from one
\emph{single} turbulent mixing layer are still far less than the observed
values, and the scenario of hundreds or thousands of TMLs per sightline might
be still needed even when these TMLs are highly supersonic.

\subsection{Convergence}
\label{sec:convergence}

\begin{figure}
    \centering
    \includegraphics[width=\columnwidth]{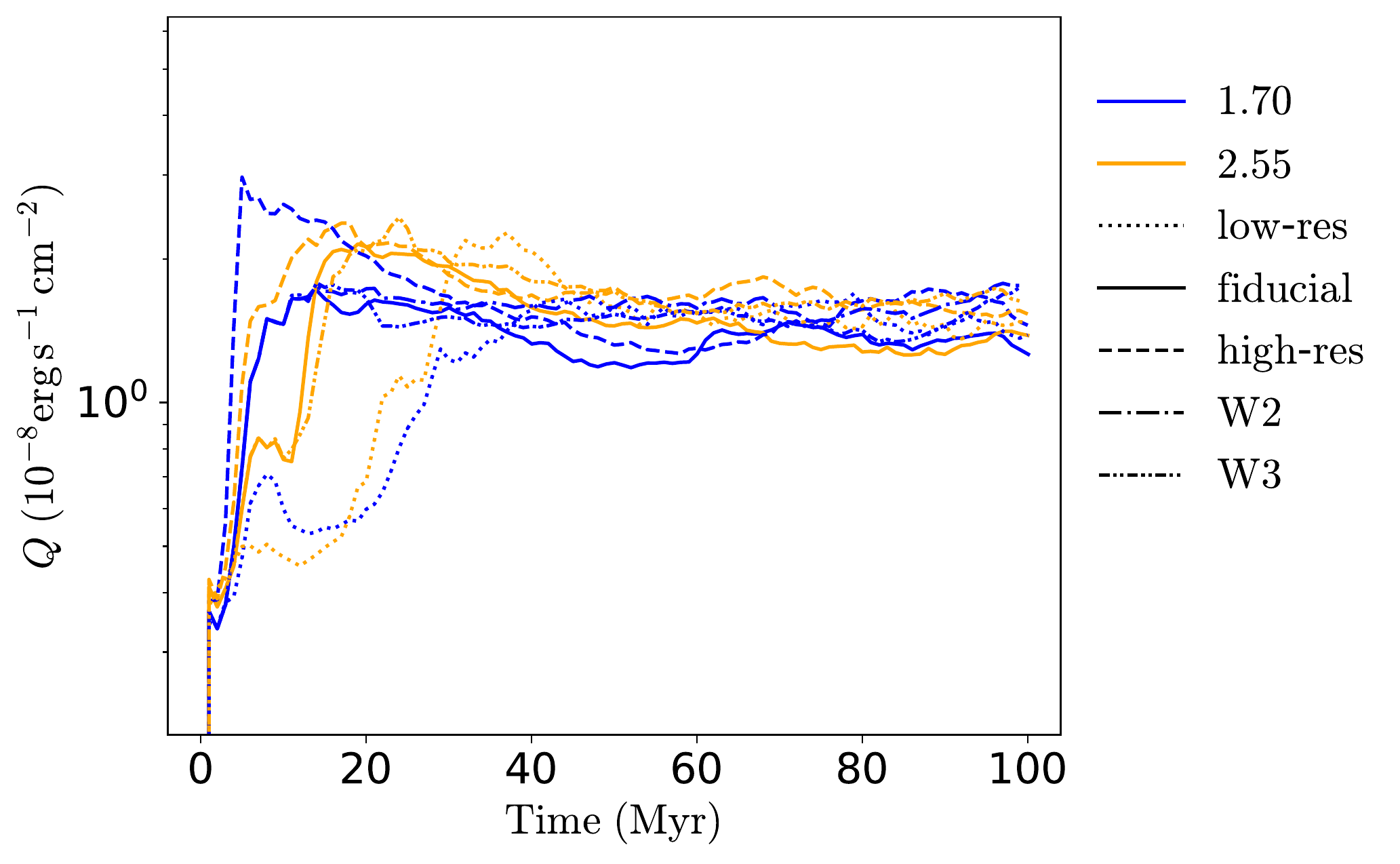}
    \caption{The time evolution of TML surface brightness $Q$ in runs with low
    (dotted), fiducial (solid) and high (dashed) resolutions, with box widths of
    100 pc (solid), 200 pc (dash-dotted), and 300 pc (dash-dot-dotted), and with
    different Mach numbers of $1.70$ (blue) and $2.55$ (orange). The
    steady-state magnitudes of $Q$ in simulations with different resolutions are
    numerically converged, even though the initial growth rates are not with
    larger rates under higher resolutions. And $Q$ exhibits negligible
    dependence on the width of the simulation box.}
    \label{fig:brightness_evo_res}
\end{figure}

\begin{figure}
    \centering
    \includegraphics[width=\columnwidth]{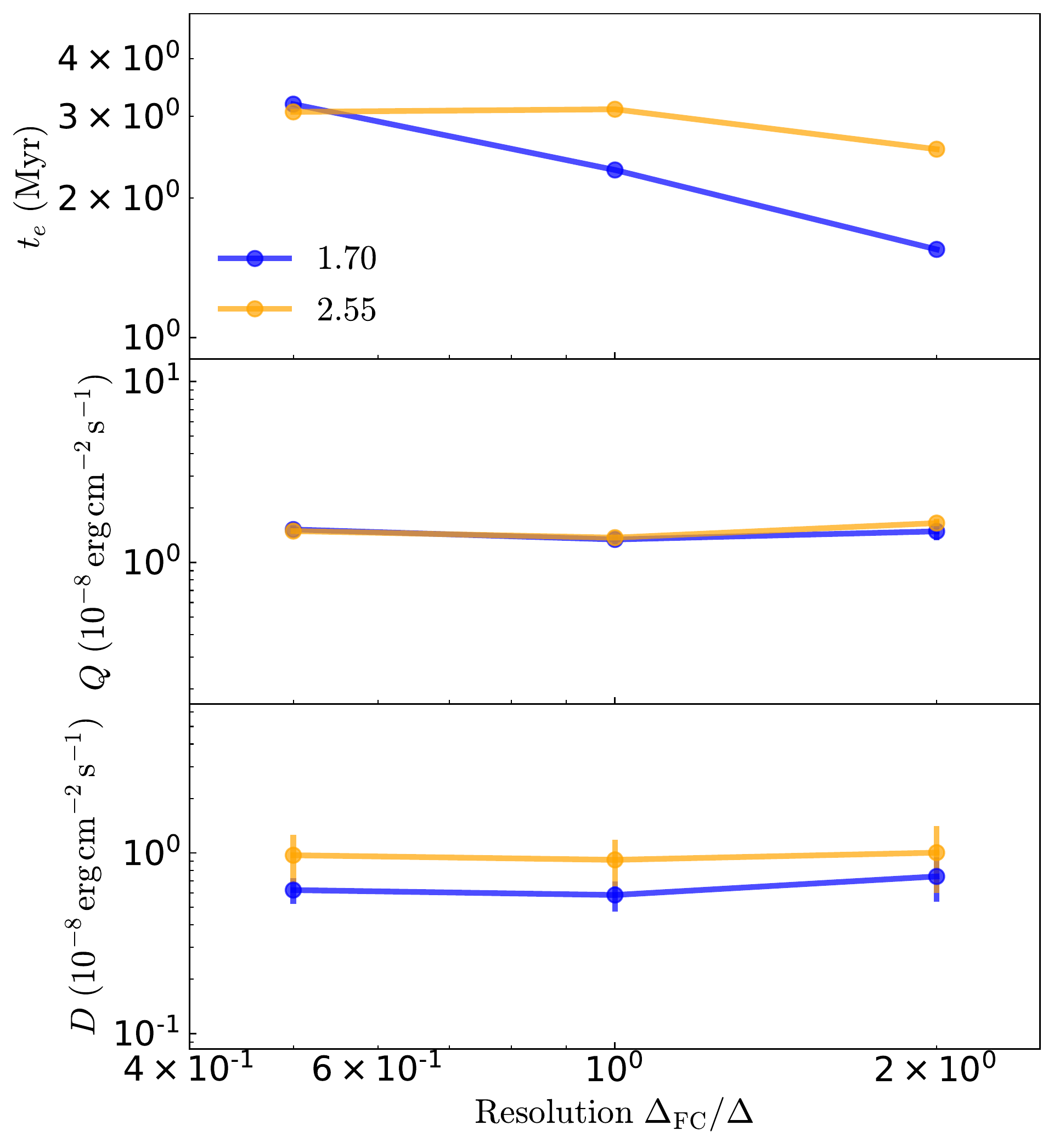}
    \caption{The e-folding time of initial turbulent velocity growth $t_e$ (top
    panel), steady-state surface brightness $Q$ (middle panel) and (column)
    dissipation rate $D$ (bottom panel) in runs with different Mach numbers of
    $1.70$ (blue) and $2.55$ (orange), plotted against resolutions used in each
    simulation. It is clearly shown that $t_e$  is resolution-dependent while
    $Q$ and $D$ are well converged.}
    \label{fig:convergence}
\end{figure}

Previous studies \citep{Ji2019,Fielding2020,Tan2021} using a similar numerical
setup have demonstrated that low-Mach number turbulent mixing layers are
generally well-converged, and here we perform convergence tests under supersonic
initial conditions, by comparing both low- and high-resolution runs for two Mach
numbers $\mathcal{M}=1.70,2.55$ against fiducial-resolution ones.
Fig.~\ref{fig:brightness_evo_res} shows the time evolution of the surface
brightness of each run. For runs with the same Mach number but different
resolutions, although the e-folding time in the growth stage does not converge,
the saturated values of surface brightness are similar, indicating the ultimate
steady state does not strongly depend on resolution at high Mach numbers. The
resolution dependence is quantitatively shown in Fig.~\ref{fig:convergence},
where the e-folding time of initial growth of the TML clearly varies with the
resolutions (top panel); however, the surface brightness and (column)
dissipation rate do converge (middle and bottom panels). We have also verified
the convergence of turbulent velocities measured from our simulations.

The non-convergence in the e-folding time is expected based on previous
discussion in \S\ref{sec:time_suppress}: since the KH instability of initial
large-wavelength perturbations is suppressed at high Mach numbers, the
later-developed KH instability has to grow from the smallest wavelengths which
are resolved by the finest grid size, so the e-folding time is
resolution-dependent. Fortunately, since the steady state is determined by
energy advection and balance (\S\ref{sec:energies}) rather than the KH
instability, the TML properties after the initial growth stage \emph{do}
converge, and our main findings focusing on steady-state TMLs are thus robust.

In addition, in order to examine the possible effect of finite domain widths
which limit the maximum allowed sizes of TML structures, we carry out three
simulations with larger domain sizes along the directions with periodic boundary
conditions (see Table~\ref{tab:special} for parameters), which are denoted by
the dash-dotted and dash-dot-dotted lines in Fig.~\ref{fig:brightness_evo_res}.
The results demonstrate that varying domain widths do not lead to qualitative
change in the saturated surface brightness with variations within a factor of
$\sim 10\%$, and simulations with fiducial domain widths are numerically robust.
The ``finite domain width effect'' is negligible due to the nature of turbulence
cascades from large to small scales: the largest eddies comparable to the domain
width at $t=0$ do not reform after turbulence is fully developed, and the finite
domain widths thus become irrelevant for the TML properties. In fact, we suspect
the ``finite domain width effect'' is even less significant at high Mach
numbers, since the KH instability of large, domain-size wavelengths are always
stabilized under supersonic conditions as previously discussed.

\section{Discussion}
\label{sec:diss}

\subsection{Caveats}
\label{sec:caveats}

In this study, we perform 3D hydrodynamic simulations to investigate properties
of the radiative turbulent mixing layers under the compressive limit. Certain
caveats apply to our study:

\begin{enumerate}
    \item \emph{The strong cooling regime} \\
    This work primarily focuses on the weak cooling regime when the cooling time
    is longer than the eddy turnover time of the turbulence ($\tau_\mathrm{turb}
    / t_\mathrm{cool} <1$).\footnote{The runs with weakest turbulence ({\tt M0.09} and {\tt M0.17}) have $\text{Da}_\text{mix} \sim 1$, while the others with higher $\mathcal{M}$ have $\text{Da}_\text{mix}< 1$ or even $\ll 1$, which means most of our simulations are in the weak cooling regime.} In this regime, the gas at intermediate temperatures
    produced by TMLs tends to be single-phase, and one-dimensional averaging of
    fluid properties (temperatures, velocities, energetics, etc.) is
    representative in our analysis. However, this might not hold in the strong
    cooling regime, when TMLs become highly multiphase on small scales and form
    fractal structures \citep{Fielding2020,Tan2021}; \citet{Tan2021} already
    demonstrated that TMLs follow different scaling relations in weak and strong
    cooling regimes. Nevertheless, the weak cooling regime studied in this work
    is typical for most galactic environments, such as the circumgalactic and
    intergalactic medium where cooling is inefficient due to low gas pressure
    and metallicity. 

    \item \emph{Magnetic fields, cosmic rays and thermal conduction} \\
    \citet{Ji2019} showed the mixing of phases can be significantly inhibited by
    magnetic tension, therefore turbulence and column densities of typical ions
    are correspondingly suppressed. It is plausible that magnetic fields are
    amplified by small-scale dynamo and suppress TMLs at high Mach numbers as
    well, while further work is required to address the impact of magnetic
    fields in the compressive limit. We also did not include cosmic rays, which
    might pressurize the gas at intermediate temperatures on both small
    \citep{wiener2017interaction} and large scales
    \citep{salem2016role,farber2018impact,butsky2018role,ji2020properties,ji2021virial,buck2020effects}.
    In addition, thermal conduction \citep{Borkowski1990,Gnat2010}, especially
    anisotropic thermal conduction could change the structure of mixing layer as
    well. Since the magnitudes of thermal conduction in galactic environments is
    less constrained, here we assume a negligible level of thermal conduction
    and rely on numerical phase mixing on grid scales. \citet{Tan2021} found the
    Field length is not necessary to be resolved for convergence, so thermal
    conduction may not be important for this particular problem, either
    physically or numerically.
 
    \item \emph{Non-equilibrium ionization and time-dependent cooling curves} \\
    We adopted the assumption of equilibrium ionization and time-independent
    cooling curves in the simulations for physical simplicity and computational
    affordability. Though non-equilibrium ionization and time-dependent cooling
    curves might alter the properties of TMLs, their effects do not make
    order-of-magnitude differences \citep{Kwak2010,Ji2019}. However, admittedly,
    these effects might be more important at high Mach numbers, since the
    dynamical time of TMLs could be even shorter than the ion recombination
    time. We leave this straightforward improvement for future work.
 
    \item \emph{Survival of cold clouds} \\
    Our conclusion on cold gas mass evolution can be directly applied to shear
    motions with large coherent length scales, such as jets and cold streams.
    Although this result is qualitatively consistent with some cloud-crushing
    simulations, this consistency needs to be viewed with caution, especially in
    the case of clouds moving supersonically, where a bow shock is produced and
    the post-shock gas returns subsonic, so supersonic TMLs becomes inapplicable
    here. In addition, since our simulations do not implement a realistic cold
    cloud, shock compression and thermalization of the cloud are not captured in
    the simulations. Therefore, we ask readers to bear this difference in mind
    when interpreting our results of cold gas survival.

\end{enumerate}

\subsection{Conclusions}
\label{conclusions}

\begin{figure}
    \centering
    \includegraphics[width=\columnwidth]{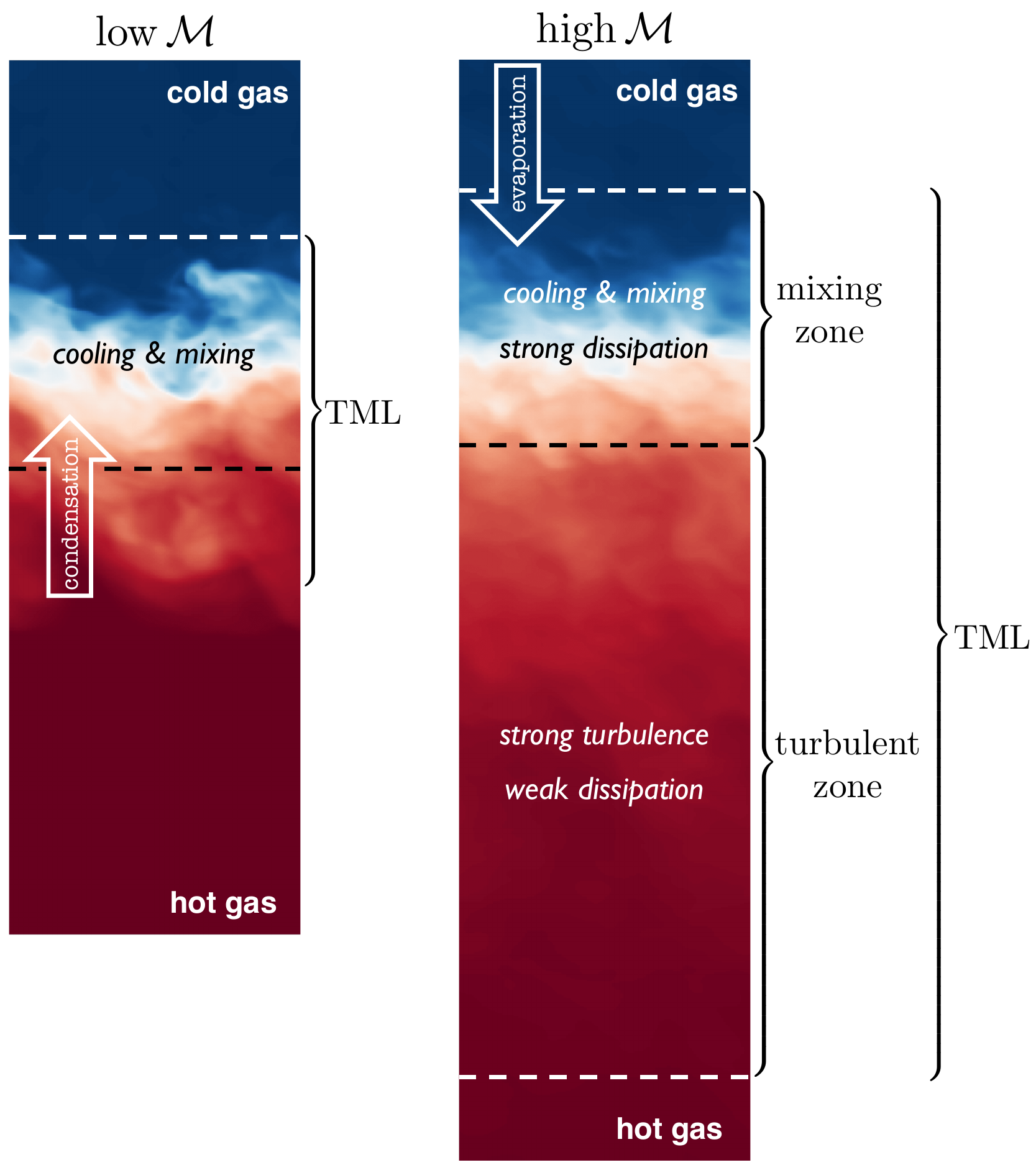}
    \caption{Schematic diagram illustrating the key features of low- and high-Mach number TMLs -- see the texts for details.}
    \label{fig:illustration}
\end{figure}

The key features of low- and high-Mach number TMLs are respectively illustrated
by Fig.~\ref{fig:illustration}, with detailed conclusions summarized as follows:

\begin{enumerate}

    \item \emph{Early-stage and long-term evolutions: from the suppression of
    the KH instability to a steady state} \\
    In the case of $\mathcal{M}>\mathcal{M}_\mathrm{crit}$, the KH instability
    of long-wavelength perturbations is initially suppressed, and TMLs can not
    develop until the jump in shear velocity profile becomes smoothed out and
    small-scale subsonic turbulence starts to emerge. The growth rate of
    turbulence in TMLs decreases with increasing $\mathcal{M}$ for
    $\mathcal{M}>\mathcal{M}_\mathrm{crit}$, while increases with greater
    $\mathcal{M}$ in the low Mach number case. When TMLs are fully developed,
    both low- and high-Mach number TMLs ultimately reach a steady state with key
    quantities (surface brightness, etc.) remaining approximately constant with
    time. At high Mach numbers, although the initial growth rate of turbulence
    is not numerically converged (since large-wavelength perturbations are
    stable to KH instability and turbulence has to grow from ``hard-to-resolve''
    small scales), TML properties in the steady state \emph{do converge}
    numerically. 

    \item \emph{Saturated surface brightness and ion column densities in
    high-Mach number TMLs} \\
    When TMLs reach a steady state, the surface brightness $Q$ and ion column
    densities $N$ ($N_\ion{O}{VI}$, etc.) in each simulation are measured, and
    our findings are as follows. At $\mathcal{M}\lesssim 1$, both $Q$ and $N$
    scale with Mach number with a power law of $\mathcal{M}^{0.5}$, while at
    $\mathcal{M}\gtrsim 1$, neither $Q$ nor $N$ increases further with greater
    $\mathcal{M}$, indicating a saturated level of surface brightness and ion
    column densities in high-Mach number TMLs, e.g., $N_\ion{O}{VI}$ saturates
    at $\sim 10^{11}\,\mathrm{cm^{-2}}$ ($0.1Z_\odot$ assumed). Therefore, the
    scenario proposed in \citet{Ji2019}, that each sightline penetrating
    hundreds or thousands TMLs is required to explain the high
    \ion{O}{VI} column densities in the circumgalactic medium\, possibly
    holds at even much higher shear velocities as well.

    \item \emph{A high-Mach number TML consisting of a turbulent zone and a
    mixing zone} \\
    We find that the shear velocity profiles of TMLs are shaped and constrained
    by local sound speeds in the zone with prominent mixing. Under supersonic
    conditions, this constraint from local sound speeds prevents both the
    turbulent velocity and size of the mixing zone from growing further with
    larger Mach numbers, while outside of the mixing zone, turbulence in the hot
    gas continues to accumulate with increasing Mach number. As a consequence, a
    high-Mach number TML develops into a \emph{turbulent zone} with large
    turbulent velocities which expands with greater $\mathcal{M}$, plus a
    \emph{mixing zone} with significant gas cooling and phase mixing. Unlike the
    turbulent zone, quantities associated with the mixing zone (e.g., surface
    brightness $Q$, ion column densities $N$) are independent of $\mathcal{M}$.
    The mixing zone has intrinsically different gas dynamics from the entire
    high-Mach number TML: turbulent velocities in the entire TML
    $u'_\mathrm{TML}$ and the mixing zone $u'_\mathrm{mix}$ follow different
    power-law relations: $u'_\mathrm{TML}\propto \mathcal{M}^{1/3}$ and
    $u'_\mathrm{mix} \propto \mathcal{M}^0$ respectively. In contrast, at low
    Mach numbers, the mixing zone is indistinguishable from the entire TML
    region, with both $u'_\mathrm{TML}$ and $u'_\mathrm{mix}$ follow the same
    power law of $\mathcal{M}^{9/10}$.

    \item \emph{Energy supplement: enthalpy consumption vs. turbulent
    dissipation}\\
    At low Mach numbers, TMLs are primarily powered by local enthalpy
    consumption against radiative cooling, and turbulent dissipation is
    negligible. However, approaching to higher Mach numbers, turbulent
    dissipation becomes increasingly dominant in energizing TMLs, while the net
    contribution from enthalpy consumption becomes less important. It also
    worths mentioning that although the turbulent zone is featured by large
    velocity dispersions, in terms of energies, it contains relatively weaker
    turbulent dissipation due to much lower densities of the hot gas; the
    turbulent dissipation actually peaks at the connecting region between the
    turbulent and mixing zones.

    \item \emph{Suppression of both inflow velocity and hot gas entrainment at
    high Mach numbers} \\
    Under the assumption of a quasi-equilibrium state where the influx of
    enthalpy and kinetic energy balances radiative cooling, the inflow velocities
    of hot gas has asymptotes of $v_\text{in}^\text{eq} \propto
    \mathcal{M}^{1/2}$ for $\mathcal{M}\to 0$ and $v_\text{in}^\text{eq} \propto
    \mathcal{M}^{-2}$ for $\mathcal{M}\to \infty$. This implies the inflow
    velocity as well as hot gas entrainment is enhanced by larger shear velocity
    at $\mathcal{M}\lesssim 1$, but is substantially \emph{suppressed} at
    $\mathcal{M}\gtrsim 1$. Moreover, the measured inflow velocity $v_\text{in}$ is
    well-predicted by $v_\text{in}^\text{eq}$ for $\mathcal{M}\lesssim 1$, but
    falls below $v_\text{in}^\text{eq}$ for $\mathcal{M}\gtrsim 1$, indicating
    high-Mach number TMLs do not reach a quasi-equilibrium state. 
    
    \item \emph{Hot gas condensation and cold gas evaporation in low- and
    high-Mach number TMLs respectively}\\
    Low-Mach number TMLs are accompanied by efficient hot gas entrainment. The
    hot gas condenses into cold phase via mixing, powering TMLs against
    radiative cooling and leading to the increase of cold gas mass. In contrast,
    the inflow velocity and hot gas entrainment are suppressed at high Mach
    numbers. Driven by strong turbulent dissipation in TMLs, cold gas evaporates
    and is ``digested'' by TMLs, with a mass loss rate increasing with higher $\mathcal{M}$. 

\end{enumerate}

\section*{Acknowledgements}
Y. Yang is grateful to his girlfriend Yuening Zhao for all her love and
encouragement. The authors thank the anonymous referee and the editor for
the constructive comments and suggestions which significantly improve this
work. The authors also thank Chad Bustard, Max Gronke, Philip
Hopkins, Linhao Ma, Peng Oh, Brent Tan, Zhiyuan Yao and Feng Yuan for helpful
discussions. The authors are supported by the Natural Science Foundation of
China (grants 12133008, 12192220, and 12192223), the science research grants
from the China Manned Space Project (No. CMS-CSST-2021-B02) and a Sherman
Fairchild Fellowship from Caltech. The simulations were performed on the High
Performance Computing Resource in the Core Facility for Advanced Research
Computing at Shanghai Astronomical Observatory, the Wheeler cluster at Caltech
and allocations FTA-Hopkins/AST20016 supported by the NSF and TACC. We have made
use of NASA's Astrophysics Data System. Data analysis and visualization are made
with {\small Python 3}, and its packages including {\small NumPy}
\citep{van2011numpy}, {\small SciPy} \citep{oliphant2007python}, {\small
Matplotlib} \citep{hunter2007matplotlib}, the {\small yt} astrophysics analysis
software suite \citep{Turk2010}, the absorption spectra tool {\small Trident}
\citep{hummels2017trident}, and the spectral simulation code {\small CLOUDY}
\citep{ferland20172017}.

\dataavailability{The data supporting the plots within this article are available on reasonable request to the corresponding author.}

\bibliographystyle{mnras}
\bibliography{refer}

\bsp
\label{lastpage}
\end{CJK}
\end{document}